\documentclass[openbib,12pt]{article}

\newcommand{\utwi}[1]{\mbox{\boldmath $ #1$}}

\usepackage{booktabs,dcolumn}
\usepackage{psfrag,graphicx}
\usepackage{eso-pic,graphicx}
\usepackage{amsmath}
\usepackage{graphicx}
\usepackage{pdflscape}
\usepackage{longtable}
\usepackage{epstopdf}
\usepackage{appendix}

\newcolumntype{z}[1]{D{.}{.}{#1}}

\newcommand{\cred}{\textcolor{red}}
\newcommand{\cb}{\textcolor{blue}}
\date{}

\begin{document}

\title{
\begin{center} {\Large \bf Semi-parametric Realized Nonlinear Conditional Autoregressive Expectile and Expected Shortfall Models} \end{center}}
\author{Chao Wang, Richard Gerlach\\
Discipline of Business Analytics, Business School, \\The University of Sydney, Australia.}

\date{} \maketitle

\begin{abstract}
\noindent
A joint conditional autoregressive expectile and Expected Shortfall framework is proposed. The framework is extended through incorporating a measurement equation which models the contemporaneous dependence between the realized measures and the latent conditional expectile. Nonlinear threshold specification is further incorporated into the proposed framework. An Bayesian Markov Chain Monte Carlo method is adapted for estimation, whose properties are assessed and compared with maximum likelihood via a simulation study. One-day-ahead VaR and ES forecasting studies, with seven market indices, provide empirical support to the proposed models.

\vspace{0.5cm}

\noindent {\it Keywords}: Expectile, Value-at-Risk, Expected Shortfall, Nonlinear, Realized Measures, Markov Chain Monte Carlo.
\end{abstract}

\newpage
\pagenumbering{arabic}

{\centering
\section{\normalsize INTRODUCTION}\label{introduction_sec}
\par
}
\noindent

Value-at-Risk (VaR) is employed by many financial institutions as an important risk management tool. Representing the market risk as one number, VaR has become a standard risk measurement metric.  However, VaR cannot measure
the expected loss for extreme (violating) returns. Expected Shortfall (ES, Artzner \emph{et al.}, 1997, 1999) calculates the average of returns on the ones being below the quantile (VaR) of its distribution, and is a more coherent measure than VaR. Thus, in recent years ES has become more widely employed for tail risk measurement and is one important change appears in the Basel Accord III (Basel Committee, 2010) which is expected to occur in the period leading up to 1st January 2019. However, there is much less existing research on modeling ES compared with VaR.

In recent two decades, the availability of high frequency data enables the calculation of various realized measures of volatility, including Realized Variance (RV): Andersen and Bollerslev (1998), Andersen \emph{et al.} (2003); and Realized Range (RR): Martens and van Dijk (2007), Christensen and Podolskij (2007), etc. Realized measures of volatility now play a key role in calculating accurate volatility estimates and
forecasts, e.g. the Realized GARCH model of Hansen, Huang and Shek (2011) and earlier work by Giot and Laurent (2004) and Clements, Galvao, and Kim (2008).

The quantile regression type model, e.g. the Conditional Autoregressive Value-at-Risk (CAViaR) model of Engle and Manganelli (2004), is a popular semi-parametric approach to forecast VaR. Gerlach, Chen and Chan (2011) generalize the CAViaR models to a fully nonlinear family. In additional, the realized measures have been employed into the quantile regression framework. \v{Z}ike\v{s} and Barun\'{i}k (2014) investigate how the conditional quantiles of future returns and volatility of financial assets vary with various realized measures. Avdulaj and Barunik (2017) explore nonlinearities in returns and propose to incorporate realized measures with the nonlinear quantile regression framework using copulas, to explain and forecast the conditional quantiles of financial returns.

However, the CAViaR type models cannot directly estimate and forecast ES. A semi-parametric model that directly estimates quantiles and expectiles, and implicitly ES, called the Conditional Autoregressive Expectile (CARE) model, is proposed by Taylor (2008). Gerlach and Chen (2016) employ the daily range into the CARE framework which is further extended into fully nonlinear family. Again, realized measures have been proved to be able to provide extra efficiency for the CARE type models (Gerlach, Walpole and Wang, 2017).

To select the appropriate expectile level, a grid search process is required for the CARE type models which is relatively computationally expensive (dependent on the model complexity and the size of the grid). As an alternative, Taylor (2017) proposes a joint ES and quantile regression framework (ES-CAViaR) which employs the Asymmetric Laplace (AL) density to build a likelihood function whose maximum likelihood estimates (MLEs)
coincide with those obtained by minimisation a joint loss function for VaR and ES. Fissler and Ziegel (2016) develop a family of joint loss functions (or ``scoring rules'') of the associated VaR and ES series that are strictly consistent for the true VaR and ES series, i.e. they are uniquely minimized by the true VaR and ES series. Under specific choices of functions in the join loss function of Fissler and Ziegel (2016), it can be shown that such loss function is exactly the same as the negative of AL log-likelihood function presented in Taylor (2017). Patton, Ziegel and Chen (2017) propose new dynamics models for VaR and ES, through adopting the generalized autoregressive score (GAS) framework (Creal, Koopman and Lucas (2013) and Harvey (2013)) and utilizing the loss functions in Fissler and Ziegel (2016).


In this paper, firstly a joint Conditional Autoregressive Expectile and Expected Shortfall (ES-CARE) framework is proposed, inspired by Engle and Manganelli (2004) and Taylor (2008, 2017). Secondly, the proposed model is extended with adding a measurement equation to incorporate realized measure to drive the tail risk dynamics (Realized-ES-CARE). Thirdly, the proposed framework is extended to nonlinear qunatile and ES autoregressive specification to model the volatility asymmetry (Realized-Threshold-ES-CARE). An adaptive Bayesian MCMC algorithm is utilised for estimation and forecasting in the proposed models. To evaluate the performance of the proposed Realized-ES-CARE and Realized-Threshold-ES-CARE models, the accuracy of the associated VaR and ES forecasts are assessed via an empirical study. Over a long forecasting period which includes 2008 GFC, results illustrate that the proposed Realized(-Threshold)-ES-CARE perform favourably, compared to Taylor's CARE and ES-CAViaR models and to a range of traditional competing models.

The paper is organized as follows:  A review of the ES-CAViaR and CARE models is conducted in Section \ref{model_review_section}. Section \ref{model_section} formalizes the proposed ES-CARE, Realized-ES-CARE and Realized-Threshold-ES-CARE models. The associated likelihood and the adaptive Bayesian MCMC algorithm for parameter estimation are presented in Section \ref{beyesian_estimation_section}.
The simulation studies are discussed in Section \ref{simulation_section}. The empirical results are presented in Section \ref{data_empirical_section}. Section \ref{conclusion_section} concludes the paper and discusses future work.

{\centering
\section{\normalsize ES-CAViaR and CARE MODELS}\label{model_review_section}
\par
}

Koenker and Machado (1999) show that the quantile regression estimator is equivalent to a maximum likelihood estimator when assuming that the data are conditionally Asymmetric Laplace (AL)
with a mode at the quantile. If $r_t$ is the data on day $t$ and $Pr(r_t < Q_t | \Omega_{t-1}) = \alpha$, then the parameters in the model for $Q_t$ can be estimated using a likelihood based on:
$$ p(r_t| \Omega_{t-1}) = \frac{\alpha (1-\alpha)}{\sigma} \exp \left( -(r_t-Q_t)(\alpha - I(r_t < Q_t)  \right)\,\, , $$
for $t=1,\ldots,n$ and where $\sigma$ is a nuisance parameter.

Taylor (2017) extends this result to incorporate the associated ES quantity into the likelihood expression, noting a link between $ES_t$ and a dynamic $\sigma_t$, resulting in the conditional density function:
\begin{eqnarray} \label{es_var_likelihood}
p(r_t| \Omega_{t-1}) = \frac{\alpha (1-\alpha)}{ES_t} \exp \left( -\frac{(r_t-Q_t)(\alpha - I(r_t < Q_t)}{\alpha ES_t}  \right)\,\, ,
\end{eqnarray}

allowing a likelihood function to be built and maximised, given model expressions for $Q_t, \text{ES}_t$. Taylor (2017) notes that the negative logarithm of the resulting likelihood function is strictly consistent
for $Q_t, \text{ES}_t$ considered jointly, e.g. it fits into the class of strictly consistent functions for VaR\&ES jointly developed by Fissler and Zeigel (2016).

Taylor (2017) incorporates two different ES components that describe the dynamics between VaR and ES and also avoid ES estimates crossing the corresponding VaR estimates, as presented in Model (\ref{es_caviar_add_model}) (ES-CAViaR-Add: ES-CAViaR with an additive VaR to ES component) and Model (\ref{es_caviar_mult_model}) (ES-CAViaR-Mult: ES-CAViaR with an multiplicative VaR to ES component):

\noindent \textbf{ES-CAViaR-Add:}

\begin{eqnarray} \label{es_caviar_add_model}
Q_{t}&=& \beta_1+ \beta_2 |r_{t-1}| + \beta_3 Q_{t-1},\\ \nonumber
 \text{ES}_t&=&Q_t-w_t, \\ \nonumber
 w_t&=&
\begin{cases}
    \gamma_0 + \gamma_1 (Q_{t-1} - r_{t-1}) + \gamma_2 w_{t-1} & \text{if } r_{t-1} \leq Q_{t-1},\\
    w_{t-1}              & \text{otherwise},
\end{cases}
\end{eqnarray}
where $\gamma_0 \ge 0, \gamma_1 \ge 0, \gamma_2 \ge 0$ are constrained in Taylor (2017), to ensure that the VaR and ES series do not cross.

\noindent \textbf{ES-CAViaR-Mult:}

\begin{eqnarray} \label{es_caviar_mult_model}
Q_{t}&=& \beta_1+ \beta_2 |r_{t-1}| + \beta_3 Q_{t-1}, \\ \nonumber
 \text{ES}_t&=& w_t Q_t, \\ \nonumber
 w_t&=& 1+\exp(\gamma_0),
\end{eqnarray}
where $\gamma_0$ is unconstrained.

\begin{figure}[htp]
     \centering
\includegraphics[width= 0.5\textwidth]{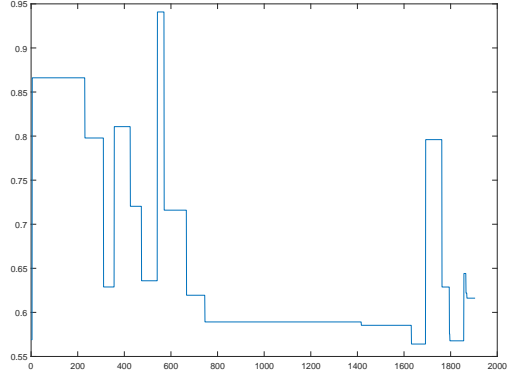}
\caption{\label{wt_example} In-sample $w_t$ plots estimated with ES-CAViaR-Add model with S\&P 500.}
\end{figure}

In addition, for $\alpha=$ 1\% the $w_t$ component for the ES-CAViaR-Add model has in-sample estimates as in Figure \ref{wt_example}.
This step function behavior occurs since $r_{t-1} \leq Q_{t-1}$ only occurs for 1\% of the observations in an accurate model. This behavior, exhibiting constant differences
between VaR and ES for long periods, and large, sustained jumps in $VaR_t-ES_t$, seems non-intuitive and potentially able to be improved. In addition, the ES-CAViaR-Mult model has a simple multiplicative VaR to ES ratio component, while there is no direct econometrics interpretation of the parameter $\gamma_0$ in the framework.

\subsection{Expectile}
Expectile is closely related to quantile. The $\tau$ level expectile $\mu_{\tau}$, defined by Aigner, Amemiya and Poirier (1976), can be estimated through minimizing the following Asymmetric Least Squares (ALS) equation (Taylor, 2008):
\begin{equation}\label{als_equation}
\sum_{t=1}^{n} |\tau-I(r_t<\mu_{\tau})|(r_t-\mu_{\tau})^2  \, ,
\end{equation}
No distributional assumption is required to estimate $\mu_{\tau}$ here.

As discussed in Section \ref{introduction_sec}, ES is defined as $\text{ES}_{\alpha}= E(Y|Y<Q_{\alpha})$, which stands for the expected
value of $Y$, conditional on the set of $Y$ that is more extreme than the $\alpha$-level quantile of Y, denoted $Q_{\alpha}$.
Newey and Powell (1987) and Taylor (2008) show that this relationship can be formulated as:
\begin{equation}\label{expectile_es_equation}
\text{ES}_{\alpha}= \left(1+\frac{\tau}{(1-2\tau)\alpha_{\tau}} \right)\mu_{\tau} \, ,
\end{equation}
where $\mu_{\tau}=Q_{\alpha}$, e.g. $\mu_{\tau}$ occurs at the quantile level $\alpha_{\tau}$ of $Y$. Thus, $\mu_{\tau}$ can be used
to estimate the $\alpha$ level quantile $Q_{\alpha}$, and then scaled to estimate the associated ES.

Taylor (2008) proposes the CARE type models which have a similar form to CAViaR models of Engle and Manganelli (2004), where lagged returns drive the expectiles,
and employed ALS for estimation. The general Symmetric Absolute Value (SAV) form of this model is:

\noindent
\textbf{CARE-SAV}:
\begin{align*}
 \mu_{t;\tau} = \beta_{1} + \beta_{2} |r_{t-1}| + \beta_{3} \mu_{t-1;\tau}
\end{align*}
\noindent
where $r_t$ is the return, $\mu_{t;\tau}$ is the $\tau$-level expectile and $r_t$ is daily return, all on day $t$. The CARE-type model produces one-step-ahead forecasts of $\mu_{t;\tau}$ (expectiles), that can be employed as VaR estimates, by an appropriate choice of $\tau$. The VaR estimates can be further scaled, using Equation (\ref{expectile_es_equation}), to produce forecasts of ES which cannot be directly calculated under the CAViaR framework.

However, the selection of appropriate expectile level $\tau$ requires a grid search, based on the violation rate (VRate, the ratio of the returns exceed the VaR estimates) or quantile loss function (Gerlach and Wang, 2016b). Specifically, for each grid value of $\tau$, the ALS estimator of the CARE equation parameters $\beta$ is found, yielding an associated VRate($\tau$). $\hat{\tau}$ is set to the grid value of $\tau$ s.t. VRate is closest to the desired $\alpha$. In the real applications, this grid search approach can be computationally expensive (dependent on the model complexity and the size of the grid), and the performance can be affected by the size and gap of the grid which is normally decided under ad-hoc approach.

\vspace{0.5cm}
{\centering
\section{\normalsize MODEL PROPOSED} \label{model_section}
}
\noindent

\subsection{ES-CARE Model}

In this paper, firstly we propose a new ES-CARE framework, inspired by Engle and Manganelli (2004) and Taylor (2008, 2017), to jointly and efficiently estimate and generate VaR \& ES forecasts. 

Given ES to $\tau$ level expectile ($\alpha$ level quantile) relation as in Equation (\ref{expectile_es_equation}), we have:
\begin{equation}\label{var_es_equation}
\mu_{\tau} = Q_{\alpha} = \frac{\text{ES}_{\alpha}}{1+\frac{\tau}{(1-2\tau)\alpha_{\tau}}}
\end{equation}

Putting Equation (\ref{var_es_equation}) into the CARE model as below:

\begin{eqnarray} \label{caviar_sav}
&& \mu_{t;\tau}= \beta_1+ \beta_2 |r_{t-1}| + \beta_3 \mu_{t-1;\tau}, \nonumber
\end{eqnarray}

we have:
\begin{eqnarray} \label{caviar_sav}
&&\frac{\text{ES}_{t;\alpha}}{1+\frac{\tau}{(1-2\tau)\alpha_{\tau}}} = \beta_1+ \beta_2 |r_{t-1}| + \beta_3 \frac{\text{ES}_{t-1;\alpha}}{1+\frac{\tau}{(1-2\tau)\alpha_{\tau}}}, \nonumber
\end{eqnarray}

thus an autoregressive framework of ES can be derived as:
\begin{eqnarray} \label{caviar_sav}
&&\text{ES}_{t;\alpha}= \beta_1 \left( 1+\frac{\tau}{(1-2\tau)\alpha_{\tau}} \right) + \beta_2 \left( 1+\frac{\tau}{(1-2\tau)\alpha_{\tau}} \right) |r_{t-1}| + \beta_3 \text{ES}_{t-1;\alpha}, \nonumber
\end{eqnarray}

Therefore, the new ES-CARE model is proposed as:

\noindent
\textbf{ES-CARE}:
\begin{eqnarray} \label{es_care_model}
\mu_{t;\tau}&=& \beta_1+ \beta_2 |r_{t-1}| + \beta_3 \mu_{t-1;\tau},\\ \nonumber
\text{ES}_{t;\alpha}&=& \beta_1 \left( 1+\frac{\tau}{(1-2\tau)\alpha} \right) + \beta_2 \left( 1+\frac{\tau}{(1-2\tau)\alpha} \right) |r_{t-1}| + \beta_3 \text{ES}_{t-1;\alpha}, \nonumber
\end{eqnarray}
subscripts $\tau$ is removed from $\alpha_{\tau}$ to simplify the notation. There are 4 parameters to be estimated in total in Model (\ref{es_care_model}): $\beta_1$, $\beta_2$, $\beta_3$, and $\tau$. $\tau$ is constraint with $[0,\alpha]$ based on its definition. Although stationarity conditions have not been theoretically considered in the literature, it is logical
that a necessary condition would be $\beta_3<1$, so that $\mu_{t;\tau}$ and $\text{ES}_{t;\alpha}$ do not diverge; but this is not a sufficient condition for stationarity. There are no other constraints for $\beta_1$, $\beta_2$ and $\beta_3$.

It is worth note that the ${1+\frac{\tau}{(1-2\tau)\alpha_{\tau}}}$ factor is equivalent to the $1+\exp(\gamma_0)$ factor in ES-CAViaR-Mult framework (Model \ref{es_caviar_mult_model}). However, the ES-CARE has a simple linear $\frac{\tau}{(1-2\tau)\alpha_{\tau}}$ function which is potentially easier to be identified with higher accuracy than the Exponential function in ES-CAViaR-Mult model. The simulation study actually lends evidence on this. In addition, the estimated $\tau$ has a direct econometrics interpretation (expectile level), and can be used to demonstrate why the ES-CARE model can be more efficient than the original CARE model. More results will be provided in later sections.

The new framework has several nice properties. Compared with the CARE model, the model can simultaneously estimate VaR (expectile), ES and the expectile level $\tau$ without any grid search, resulting in significantly speed up estimation process. Further, the $\tau$ is estimated under a VaR and ES join loss function, e.g. Equation (\ref{es_caviar_like_equation}), so it is a more statistical estimation procedure compared with the existing ad-hoc grid search, which can potentially improve the VaR and ES estimation and forecasting accuracy. More evidence will be provided in the later sections on the improved $\tau$, VaR and ES results. In addition, compared with the ES-CAViaR-Add model in Taylor (2017), the ES-CARE framework has a more parsimonious and dynamic ES component, which can potentially tackle the challenges presented in Figure \ref{wt_example}.

Also, the ES and VaR (expectile) are guaranteed to be not cross with each other based on the above derivations. Later on, we will provide more empirical evidence on the improved VaR and ES forecasting performance with ES-CARE, compared with CARE and ES-CAViaR.

Lastly, the ES-CARE model employs autoregressive specifications for both Expectile and ES, which enables the development of fully nonlinear threshold expectile and ES autoregressive dynamics.

\subsection{Realized(-Threshold)-ES-CARE Models}

The Realized-GARCH (Re-GARCH) framework is proposed in Hansen, Huang and Shek (2012). Compared to the conventional GARCH model, the Re-GARCH employs a measurement equation, which captures the contemporaneous relation between unobserved volatility and a realized measure. The superiority of Re-GARCH compared to GARCH and GARCH-X is well demonstrated, e.g. see Hansen, Huang and Shek (2012), Watanabe (2012) and Gerlach and Wang (2016a).

The Realized-ES-CARE (Re-ES-CARE) framework is proposed as below, through adding a measurement equation which models the relation between expectile and a realized measure into the ES-CARE framework.

\noindent
\textbf{Re-ES-CARE:}
\begin{eqnarray} \label{re_es_care}
\mu_{t;\tau}&=& \beta_1+ \beta_2 X_{t-1} + \beta_3 \mu_{t-1;\tau},\\ \nonumber
\text{ES}_{t;\alpha}&=& \beta_1 \left( 1+\frac{\tau}{(1-2\tau)\alpha} \right) + \beta_2 \left( 1+\frac{\tau}{(1-2\tau)\alpha} \right) X_{t-1} + \beta_3 \text{ES}_{t-1;\alpha}, \\ \nonumber
 X_t&=& \xi+\phi |\mu_{t;\tau}|+ \delta_1 \epsilon_t + \delta_2 (\epsilon_t^2-E(\epsilon^2)) + u_t \, , \\ \nonumber
\end{eqnarray}
where $X_t$ is a realized measure observed on day $t$, details to be discussed in Section \ref{data_empirical_section}.  The measurement equation here is of a standard time series form, e.g. $E(u_t) = 0$, thus the standard setting and choice of $u_t \stackrel{\rm i.i.d.} {\sim} N(0,\sigma_{u}^2)$ is made for the measurement error.

It is important to note that neither the likelihood for the ES-CARE models nor for the Realized-ES-CARE models is a parametric likelihood or leads to a parametric MLE. The likelihood assumes a given
value for $\alpha$ during estimation, thus directly targeting a specific expectile (quantile) of the conditional return distribution, without assuming it has a specific distributional form.

Compared to the ES-CAViaR or Realized-GARCH models which have only one return-related ``error'', there are two return-related ``error'' series in the proposed Realized-ES-CARE type models: one is the $z_t = r_t -\mu_{t;\tau}$, which is assumed to follow an asymmetric Laplace distribution with time varying scale, so that likelihood can be constructed based on this AL density to jointly estimate the conditional VaR and conditional ES. However, the framework does not rely on an AL or any distribution assumption for the returns. The other one is $\epsilon_t= \frac{r_t} {\mu_{t;\tau}}$, that appears in the measurement equation and is employed to capture the well known leverage effect.
Again, if $\mu_{t;\tau}$ is a multiple of $\sqrt{h_t}$ then, we will have $E(\epsilon_t)=0$, as usual, but to keep a
zero mean asymmetry term $(\epsilon_t^2-E(\epsilon^2))$, we need to know $$ E(\epsilon^2) = E\left(\frac{r_t^2}{\mu_{t;\tau}^2}\right). $$

This second moment information is not included in Realized-ES-CARE framework. Thus, we substitute it with an empirical estimate
$E(\epsilon^2) \approx \bar{\epsilon^2}$, being the sample mean of the squared multiplicative errors. We note
that $E(\epsilon_t^2-\bar{\epsilon^2})= 0$ is preserved if $\bar{\epsilon^2}$ is an unbiased estimate. Therefore, the term
$\delta_1 \epsilon_t + \delta_2 (\epsilon_t^2-\bar{\epsilon^2})$ still generates an asymmetric response in volatility to return shocks.
Further, the sign of $\delta_1$ is expected to be opposite to that from an Realized-GARCH model, since the expectile $\mu_{t;\tau}$ is negative for
the lower quantile levels , e.g. $\alpha=1\%$, considered in the paper.

Motivated by the nonlinear quantile dynamics in Gerlach, Chen and Chan (2011), the Realized-ES-CARE framework is further extended to the threshold nonlinear specifications. In addition to the nonlinear expectile (VaR) component, a non-linear ES autoregressive component is incorporated. This is benefited from the proposed ES-CARE framework which directly incorporates an ES autoregressive component. The model is named as Realized-Threshold-ES-CARE (Re-Threshold-ES-CARE):

\textbf{Re-Threshold-ES-CARE:}
\begin{eqnarray} \label{re_es_care_t}
\mu_{t;\tau}&=&
    \begin{cases}
        \beta_{1}+ \beta_{2} X_{t-1} + \beta_{3} \mu_{t-1;\tau} , &  z_{t-1} \leq c, \\
        \beta_{4}+ \beta_{5} X_{t-1} + \beta_{6} \mu_{t-1;\tau} , &  z_{t-1} > c,
    \end{cases} \\ \nonumber
\text{ES}_{t;\alpha}&=&
    \begin{cases}
        \beta_{1} \left( 1+\frac{\tau}{(1-2\tau)\alpha} \right) + \beta_{2} \left( 1+\frac{\tau}{(1-2\tau)\alpha} \right) X_{t-1} + \beta_{3} \text{ES}_{t-1;\alpha} , &  z_{t-1} \leq c, \\
        \beta_{4} \left( 1+\frac{\tau}{(1-2\tau)\alpha} \right) + \beta_{5} \left( 1+\frac{\tau}{(1-2\tau)\alpha} \right) X_{t-1} + \beta_{6} \text{ES}_{t-1;\alpha}  , &  z_{t-1} > c,
    \end{cases} \\ \nonumber
X_t&=& \xi+\phi |\mu_{t;\tau}|+ \tau_1 \epsilon_t + \tau_2 (\epsilon_t^2-E(\epsilon^2)) + u_t \, , \\ \nonumber
\end{eqnarray}

$z_{t}$ is a threshold variable and chosen to be self-exciting, e.g., $z_{t}= r_{t}$, and $c$ is the threshold value and set as 0 in our paper.

{\centering
\section{\normalsize LIKELIHOOD AND BAYESIAN ESTIMATION} \label{beyesian_estimation_section}
\par
}
\noindent
\subsection{ES-CARE Likelihood Function with AL}\label{es_caviar_likelihood_section}

Taylor (2017) extended the Koenker and Machado (1999) result to incorporate the ES in the equivalent likelihood function which is given in Equation (\ref{es_caviar_like_equation}). Note here $\mu_{t;\tau}= Q_t$ as discussed in Section \ref{model_section}.

\begin{eqnarray}\label{es_caviar_like_equation}
\ell(\mathbf{r};\mathbf{\theta})= \sum_{t=1}^{n} \left( \text{log}  \frac{(\alpha-1)}{\text{ES}_t} + {\frac{(r_t- \mu_{t;\tau})(\alpha-I(r_t\leq \mu_{t;\tau}))}{\alpha \text{ES}_t}} \right).
\end{eqnarray}

\subsection{Realized(-Threshold)-ES-CARE Log Likelihood}\label{re_es_caviar_likelihood_section}
Because the Realized-ES-CARE framework has a measurement equation, with $u_t \stackrel{\rm i.i.d.} {\sim} N(0,\sigma_{u}^2)$, the full log-likelihood function for Realized-ES-CARE  is the sum of the log-likelihood $\ell (\mathbf{r};\mathbf{\theta})$ for the expectile and ES equation (as in Equation \ref{es_caviar_like_equation}) and the log-likelihood $\ell (\mathbf{x}|\mathbf{r};\mathbf{\theta})$ from the measurement equation:

\begin{eqnarray}\label{re_es_care_like_equation}
&\ell(\mathbf{r},\mathbf{X};\mathbf{\theta})= \ell(\mathbf{r};\mathbf{\theta})+ \ell(\mathbf{X}|\mathbf{r};\mathbf{\theta})=\\ \nonumber
& \underbrace{\sum_{t=1}^{n} \left( \text{log}  \frac{(\alpha-1)}{\text{ES}_t} + {\frac{(r_t-\mu_{t-1;\tau})(\alpha-I(r_t\leq \mu_{t-1;\tau}))}{\alpha \text{ES}_t}} \right) }_{\ell (\mathbf{r};\mathbf{\theta})}\\ \nonumber
& \underbrace{-\frac {1}{2} \sum_{t=1}^{n} \big( \text{log} (2 \pi)+ \text{log}(\sigma_{u}^2)+
   u_t^2/\sigma_{u}^2 \big)}_{\ell (\mathbf{X}|\mathbf{r};\mathbf{\theta})} ,\\ \nonumber
\end{eqnarray}
where $u_t= X_t- \xi - \phi |Q_{t}| - \tau_{1} \epsilon_{t} - \tau_{2} (\epsilon_{t}^2-\bar{\epsilon_{t}^2})$, $t=1,\ldots,n$.

Further, the log-likelihood function of the nonlinear Realized-Threshold-ES-CARE is the same as the Realized-ES-CARE model, except the changing nonlinear dynamics of the expectile (VaR) and ES.

\subsection{Maximum Likelihood Estimation}

We have incorporated a three step maximum Likelihood (ML) approach for the proposed Realized(-Threshold)-ES-CARE models.

In the first step, for Realized-ES-CARE, the expectile equation parameters ($\beta_1,\beta_2,\beta_3$) are estimated separately by optimizing pseudo-likelihood for a expectile regression. For Realized-Threshold-ES-CARE, a threshold expectile regression model (Gerlach and Chen, 2016) is estimated separately to get the threshold expectile equation parameters ($\beta_1$ to $\beta_6$).

In the second step, multiple starting values for the measurement equation parameters ($\xi, \phi, \tau_{1}, \tau_{2}, \sigma_{u}$) and $\tau$ are randomly sampled: 10,000 random candidate starting vectors are
used.

Finally, the estimates for (threshold) quantile equation parameters in the first step are combined with the randomly sampled candidates in the second step. The parameter set that maximizes
the log-likelihood function (\ref{re_es_care_like_equation}) is selected as the starting values for the constrained optimization routine $fmincon$ in Matlab, to generate the final ML estimates.

\subsection{Bayesian Estimation}

Motivated by the favourable estimation results for CAViaR (Gerlach, Chen and Chan, 2011) and CARE-X models (Gerlach and Chen, 2016), compared to the associated MLEs, a Bayesian
estimator is also considered.

Given a likelihood function, and the specification of a prior distribution, Bayesian algorithms can be employed to estimate the parameters of Realized-ES-CARE and Realized-Threshold-ES-CARE models.
An adaptive MCMC method, adopted from that in Gerlach and Wang (2016a) and Chen \emph{et al.} (2017) is employed in this case. Three parameter blocks were employed in the MCMC simulation:
$\utwi{\theta_1}=(\beta_1,\beta_2,\beta_3, \phi)$ for Realized-ES-CARE and $\utwi{\theta_1}=(\beta_1,\beta_2,\beta_3,\beta_4,\beta_5,\beta_6,\phi)$ for Realized-Threshold-ES-CARE,  $\utwi{\theta_2}=(\tau)$, $\utwi{\theta_3}=(\xi, \delta_{1}, \delta_{2}, \sigma_{u})$, via the motivation that parameters within the
same block are likely to be more strongly correlated in the posterior, than those between blocks, allowing faster mixing of the chain (e.g. see Damien \emph{et al.}, 2013). Priors are chosen to be uninformative over the possible stationarity and positivity regions, e.g. $\pi(\utwi{\theta})\propto I(A)$, which is a flat prior for $\utwi{\theta}$ over the region $A$.

In "burn-in" period, the "epoch" method in Chen \emph{et al.} (2017) is employed. For the initial "epoch", a Metropolis algorithm (Metropolis \emph{et al.}, 1953) employing a
mixture of 3 Gaussian proposal distributions, with a random walk mean vector, is utilised for each block of parameters. The proposal var-cov matrix of each block in each mixture element is
$C_i \Sigma$, where $C_1 =1; C_2 =100; C_3 =0.01$ (allowing both very big ($i=2$) and very small ($i=3$) jumps), with $\Sigma$ initially set to $\frac{2.38}{\sqrt{(d_i)}}I_{d_i}$. Here $d_i$ is the dimension of
block ($i$); $I_{d_i}$ is the identity matrix of dimension $d_i$. The covariance matrix is subsequently tuned, aiming towards a target acceptance rate of $23.4\%$ (if $d_i>4$, or $35\%$ if $2 \le d_i \le 4$, or $44\%$
if $d_i=1$), as standard, via the algorithm of Roberts, Gelman and Gilks (1997).

In order to enhance the convergence of the chain, at the end of 1st epoch, e.g. 20,000 iterations, the covariance matrix for each parameter block
is calculated, after discarding the first e.g. 2,000 iterations, which is used in the proposal distribution in the next epoch (of e.g. 20,000 iterations).
After each epoch, the standard deviations of each parameter chain in that epoch are calculated and are collectively compared to the standard deviations from the previous epoch. This
process is continued until the mean absolute percentage change over the standard deviations of parameters is less than a pre-specified threshold (10\% is employed in the paper).
In the empirical study, on average it takes 6-10 epochs to observe an absolute percentage change lower than 10\%; thus, the chains are run in
total for 120,000-200,000 iterations as a burn-in period, in the empirical parts of this paper. A final epoch of 12,000 iterates is run with an "independent" Metropolis-Hastings algorithm,
employing a mixture of three Gaussian proposal distributions for each block. The mean vector
for each block is set as the sample mean vector of the last epoch iterates (after discarding the first 2,000 iterates) for that block. The proposal var-cov matrix in each element is $C_i \Sigma$, where
$C_1 =1;C_2 =100;C_3 =0.01$ and $\Sigma$ is the sample covariance matrix of the last epoch iterates for that block (after discarding the first 2,000 iterates). This final epoch is employed as the
sample period, where all estimation and inference (and forecasting) is done via the posterior mean.

{\centering
\section{\normalsize Realized(-Threshold)-ES-CARE SIMULATION STUDY}\label{simulation_section}
\par
}
\noindent
Simulation studies are conducted to compare the properties and performance of the Bayesian method and MLE for Realized-Threshold-ES-CARE type models, with respect to parameter estimation
and one-step-ahead VaR and ES forecasting accuracy. To compare the bias and precision performance of the MCMC and ML methods, both the mean and Root Mean Square Error (RMSE) values are calculated over the replicated datasets.

1000 replicated return series are simulated from the following specific square root Realized(-Threshold)-GARCH model, specified as Simulation Models 1 \& 2. $n=1900$ is approximately the average in-sample (fixed) size for the empirical study across 7 indices, details as in Table \ref{var_fore_table}. To match up with the forecasting study and to find properties for the estimators in a similar situation, $n=1900$ is selected as the sample size in the simulation study.

\textbf{Simulation Model 1}
\begin{eqnarray} \label{re_garch_simu}
r_t&=& \sqrt{h_t} \varepsilon_t^{*}, \\ \nonumber
\sqrt{h_t}&=& 0.02 + 0.10 X_{t-1}+ 0.85 \sqrt{h_{t-1}},  \\ \nonumber
X_t&=& 0.1+0.9 \sqrt{h_t}-0.02 \varepsilon_t^{*}  + 0.02 (\varepsilon_t^{*2}-1) +u_t, \\  \nonumber
&&\varepsilon_t^{*} \stackrel{\rm i.i.d.} {\sim} N(0,1), u_t \stackrel{\rm i.i.d.} {\sim} N(0,0.3^2).\\  \nonumber
\end{eqnarray}

\textbf{Simulation Model 2}
\begin{eqnarray} \label{re_t_garch_simu}
r_t&=& \sqrt{h_t} \varepsilon_t^{*}, \\ \nonumber
\sqrt{h_t}&=&
\begin{cases}
        0.05 + 0.20 X_{t-1}+ 0.80 \sqrt{h_{t-1}}, &  r_{t-1} \leq 0, \\
        0.10 + 0.10 X_{t-1}+ 0.75 \sqrt{h_{t-1}}, &  r_{t-1} > 0,
\end{cases} \\ \nonumber
X_t&=& 0.1+0.9 \sqrt{h_t}-0.02 \varepsilon_t^{*}  + 0.02 (\varepsilon_t^{*2}-1) +u_t, \\  \nonumber
&&\varepsilon_t^{*} \stackrel{\rm i.i.d.} {\sim} N(0,1), u_t \stackrel{\rm i.i.d.} {\sim} N(0,0.3^2).\\  \nonumber
\end{eqnarray}

In order to calculate the corresponding Realized-ES-CARE true parameter values, a mapping from the Simulation Model 1 to the Realized-ES-CARE is required.

Further, given $\Phi^{-1}(\alpha)$ as the standard Gaussian inverse cdf, we have $\mu_{t;\tau}=Q_{t}=\sqrt{h_t} \Phi^{-1}(\alpha)$, then
$\sqrt{h_t} =\frac{\mu_{t;\tau}} {\Phi^{-1}(\alpha)} $. Then, with $\varepsilon_t^{*} \stackrel{\rm i.i.d.} {\sim} N(0,1)$, we have $\epsilon_{t}= \frac{r_t} {\mu_{t;\tau}} = \frac{r_t} {\sqrt{h_t} \Phi^{-1}(\alpha)} = \frac{\varepsilon_t^{*}} {\Phi^{-1}(\alpha)}$. Substituting these back into the Simulation Model 1, the corresponding Realized-ES-CARE specification can be written as:

\begin{eqnarray}
\mu_{t;\tau}&=& 0.02 \Phi^{-1}(\alpha)+ 0.10 \Phi^{-1}(\alpha) X_{t-1}+ 0.85\mu_{t-1;\tau},\\ \nonumber
\text{ES}_{t;\alpha}&=& 0.02 \Phi^{-1}(\alpha) \left( 1+\frac{\tau}{(1-2\tau)\alpha} \right) + 0.10 \Phi^{-1}(\alpha) \left( 1+\frac{\tau}{(1-2\tau)\alpha} \right) |r_{t-1}| + 0.85 \text{ES}_{t-1;\alpha}, \\ \nonumber
X_t &=&  0.1-\frac{0.9} {\Phi^{-1}(\alpha)} |\mu_{t;\tau}| -0.02\Phi^{-1}(\alpha)\epsilon_{t} + 0.02 \Phi^{-1}(\alpha)^2 (\epsilon_{t}^2-  \frac{1} {\Phi^{-1}(\alpha)^2} ) + u_t, \\  \nonumber
\end{eqnarray}
allowing true parameter values to be calculated or read off. These true values are presented in Table \ref{simu_table_1}. Similarly, the parameter true values of the Realized-Threshold-ES-CARE model corresponding to Simulation Model 2 are derived similarly and presented in Table \ref{simu_table_2}.

The true value of $\tau$ parameter can be calculated as well. The true in-sample and one-step-ahead $\alpha$ level VaR and ES forecast can be exactly calculated for each data set; i.e. $\text{VaR}_{t;\alpha}=\mu_{t;\tau}=Q_{t;\alpha}= \sqrt{h_{t}}\Phi^{-1}(\alpha)$,
and $\text{ES}_{t;\alpha}= -\sqrt{h_{t}} \frac{\phi(\Phi^{-1}(\alpha))}{\alpha} $, where $\phi()$ is standard Normal pdf. Via the one-to-one relationship between VaR and ES presented in Equation (\ref{expectile_es_equation}), the true value of $\tau$ in this model can be exactly calculated: 0.001461.

VaR forecast is $\text{VaR}_{n+1}= \sqrt{h}_{n+1}\Phi^{-1}(\alpha)$, and the corresponding true ES forecast is $\text{ES}_{n+1}= -\sqrt{h}_{n+1}\frac{\phi\left(\Phi^{-1}(\alpha)\right)}{\alpha}$, where $\phi$ is the standard Gaussian pdf, which are calculated for each dataset; the averages of these, over
the 1000 datasets, are given as $\text{VaR}_{n+1}$ and $\text{ES}_{n+1}$ in the "True" column of Table \ref{simu_table_1} and \ref{simu_table_2}.

The Realized-ES-CARE and Realized-Threshold-ES-CARE models are fit to the 1000 datasets generated by Simulation Models 1 and 2 respectively, once using the adaptive MCMC method and once using the ML estimator.

Estimation results of Realized-ES-CARE are summarized in Table \ref{simu_table_1}, where boxes indicate the preferred model in terms of minimum
bias (Mean closest to Ture) and maximum precision (minimum RMSE). First, both MCMC and ML generate relatively accurate parameter estimates and VaR \& ES forecasts in this case, which proves the validity of both methods
as discussed in Section \ref{beyesian_estimation_section}. For all 9 parameters and both VaR \& ES forecasts the bias results clearly favour the MCMC estimator compared to the ML. Further, the precision is higher for the MCMC method for 7 out of 9 parameters and for both VaR and ES forecasts. It is worth noting that the proposed framework can generate very close to True $\tau$ estimates, which proves the validating of the proposed framework.

As discussed in Section \ref{model_section}, the ES-CARE model has a simple linear $\frac{\tau}{(1-2\tau)\alpha_{\tau}}$ function which is potentially easier to be estimated with higher accuracy than the Exponential function in ES-CAViaR-Mult model. As can be seen, the RMSE values for the $\tau$ are quite small for both methods and are much smaller than that of the $\gamma_0$ of ES-CAViaR-Mult (simulation results not shown here). In the measurement equation, the MCMC generates clearly better estimation results for $\xi$ and $\phi$ which are known to be the two most important parameters in the realized GARCH framework.

With respect to the Realized-Threshold-ES-CARE estimation, as in Table \ref{simu_table_2}, MCMC still demonstrates it advantageous compared to ML. Accurate parameter estimates and VaR \& ES forecasting results are produced by both adaptive MCMC and ML. However, compared with the RMSE results for VaR and ES forecasts in Table \ref{simu_table_1}, increased RMSE values are observed, which is due to the challenge of estimating a more complex framework. With respect to bias results, MCMC is favoured by 5 parameters and by both the VaR and ES tail risk forecasts. Regarding precision, MCMC produces lower RMSE for 10 out 12 parameters and both VaR and ES forecasts.

\begin{table}[!ht]
\begin{center}
\caption{\label{simu_table_1} \small Summary statistics for the two estimators of the Realized-ES-CARE model, with data generated from Simulation Model 1.}\tabcolsep=10pt
\begin{tabular}{lcccccccc} \hline
$n=1900$               &             & \multicolumn{2}{c}{MCMC}      &  \multicolumn{2}{c}{ML}   \\
Parameter              &True         &Mean           &  RMSE         &Mean           & RMSE    \\ \hline
$\beta_1$         &    -0.0465 & 	\fbox{-0.0674}	& \fbox{0.1245}& -0.0714	& 0.1574   \\
$\beta_2$         &       -0.2326 	& \fbox{-0.2492}& 	\fbox{0.0904}& -0.2495& 0.0959\\
$\beta_3$          &     0.8500 	& \fbox{0.8255}	& \fbox{0.1061} & 0.8222& 0.1374   \\
$\tau$          &    0.001461 	& \fbox{0.001363} &	 \fbox{0.000311} &	 0.001348 	 &0.000322       \\
$\xi$      &   0.1000 & \fbox{0.1807}	& \fbox{0.1693}& 0.2027	& 0.3503  \\
$\phi$              &    0.3869 & 	\fbox{0.3394}	& \fbox{0.1278}& 	0.3220& 	0.2646 \\
$\delta_{1}$               &   0.0465&  	\fbox{0.0411}& 0.0168& 	0.0410& 	\fbox{0.0167}  \\
$\delta_{2}$               &   0.1082 & 	\fbox{0.0961}& 	\fbox{0.0289}	& \fbox{0.0961}& \fbox{0.0289}\\
$\sigma_{u}$           &  0.3000 	& \fbox{0.2801}& 	\fbox{0.0204}& 0.2797& 0.0208 \\
$\text{VaR}_{n+1}$     &   -1.2523& 	\fbox{-1.2499}& \fbox{0.0706}& 	-1.2497	& 0.0747\\
$\text{ES}_{n+1}$     &    -1.4349 & \fbox{-1.4203}& 	\fbox{0.0858}& 	-1.4182& 0.0896\\
\hline
\end{tabular}
\end{center}
\emph{Note}:\small A box indicates the favored estimators, based on mean and RMSE.
\end{table}

\begin{table}[!ht]
\begin{center}
\caption{\label{simu_table_2} \small Summary statistics for the two estimators of the Realized-Threshold-ES-CARE model, with data generated from Simulation Model 2.}\tabcolsep=10pt
\begin{tabular}{lcccccccc} \hline
$n=1900$               &             & \multicolumn{2}{c}{MCMC}      &  \multicolumn{2}{c}{ML}   \\
Parameter              &True         &Mean           &  RMSE         &Mean           & RMSE    \\ \hline
$\beta_1$         &  -0.1163 &	-0.1483	& \fbox{0.1573}	&\fbox{-0.0992}&	0.2033   	\\
$\beta_2$         &  -0.4653 &	-0.5096&	\fbox{0.1720}&	\fbox{-0.4961}&	0.1727	\\
$\beta_3$          &   0.8000 &	0.7688&	\fbox{0.0776}	&\fbox{0.7952}	&0.0922  \\
$\beta_4$         &  -0.2326 &	-0.2077&	\fbox{0.1521}&	\fbox{-0.2255} &	0.1816\\
$\beta_5$         & -0.2326& 	-0.2610	& \fbox{0.1452}	&\fbox{-0.2527}&	0.1476 \\
$\beta_6$          &    0.7500& 	\fbox{0.7460}&	\fbox{0.0871}&	0.7420&	0.0974\\
$\tau$            &   0.001461 &	 \fbox{0.001318}& 	 \fbox{0.000320} & 	 0.001296& 	 0.000345\\
$\xi$              &   0.1000 &	\fbox{0.1416}	& \fbox{0.2127}	&0.1465	&0.2130\\
$\phi$             &  0.3869 &	\fbox{0.3714}	&\fbox{0.0937}&	0.3696	&0.0940 \\
$\delta_{1}$        &  0.0465 &	\fbox{0.0455}	&0.0158	&0.0454	&\fbox{0.0157}   \\
$\delta_{2}$        &     0.1082& 	0.1094&	0.0278&	\fbox{0.1092}&	\fbox{0.0277} \\
$\sigma_{u}$        &  0.3000& 	\fbox{0.2999}	&\fbox{0.0051}	&0.2994	& \fbox{0.0051}\\
$\text{VaR}_{n+1}$  &  -2.3443& 	\fbox{-2.3405}	& \fbox{0.1496} &	-2.3074&0.2384\\
$\text{ES}_{n+1}$   & 	-2.6863& 	\fbox{-2.6495}	&\fbox{0.1825}	&-2.6068&	0.2822 \\
\hline
\end{tabular}
\end{center}
\emph{Note}:\small A box indicates the favored estimators, based on mean and RMSE.
\end{table}

{\centering
\section{\normalsize DATA and EMPIRICAL STUDY}\label{data_empirical_section}
\par
}
\subsection{Realized Measures}

Various realized measures, including realized variance (RV) and realized range (RR) are incorporated in the proposed Realized-ES-CARE and Realized-Threshold-ES-CARE type models.

To reduce the effect of microstructure noise of realized measures, Martens and van Dijk (2007) present a scaling process which is inspired by the fact that the daily squared return and range are less affected by microstructure noise than their high frequency counterparts. Therefore, the process can be used to smooth and scale RV and RR, creating less microstructure sensitive measures.

Further, Zhang, Mykland and A\"{i}t-Sahalia (2005) propose a sub-sampling process to deal with micro-structure effects for realized variance (SSRV). The sub-sampling process is applied to RR in Gerlach and Wang (2016b).
The properties of the sub-sampled RR, compared to those of other realized measures, are assessed via simulation under three scenarios in Gerlach and Wang (2016b).

The scaled RV (ScRV), Scaled RR (ScRR), sub-sampled RV (SSRV) and sub-sampled RR (SSRV) are also employed and tested in the proposed frameworks. For example, Realized-ES-CARE-RV represents Realized-ES-CARE framework employing RV, and Realized-Threshold-ES-CARE-RR represents Realized-Threshold-ES-CARE framework employing RR.

\subsection{Data Description}
Daily and high frequency data, observed at 1-minute and 5-minute frequency within trading hours, including open, high, low and closing prices, are downloaded from
Thomson Reuters Tick History. Data are collected for 7 market indices: S\&P500, NASDAQ (both US), Hang Seng (Hong Kong), FTSE 100 (UK),
DAX (Germany), SMI (Swiss) and ASX200 (Australia). The time period is Jan 2000 to June 2016.

The daily return and the daily RV, RR, Scaled RV and Scaled RR measures (using 5 minute data) are calculated; 1-minute data are employed to produce daily sub-sampled RV and sub-sampled RR measures; $q=66$ is employed for the scaling process as suggested in , e.g. around 3 months as suggested in Martens and van Dijk (2007).

\subsection{In-sample Parameter Estimates} \label{parameter_estimates_section}

Before presenting the forecasting results, the parameter estimates from Realized-Threshold-ES-CARE models are shown for S\&P 500. Table \ref{para_est_table} presents the parameter estimates for the 1st forecasting step (using 1st in-sample data set).

First, the parameter estimates are consistent with the results in Table II of Hansen, Huang and Shek (2012), after mapping between Realized-Threshold-ES-CARE and Realized-GARCH model as discussed in Section \ref{simulation_section}. For example, the $\varphi$ estimates in the measurement equation should be in general close to 1 in the Realized-GARCH framework.  Dividing 1 by $\Phi^{-1}(\alpha)$ (assuming Normal error distribution) and taking negative of it (as $|\mu_{t;\tau}|$ employed), we have 0.4299 which is close to real the $\varphi$ estimates in Table \ref{para_est_table}.

Second, we can see that the absolute values of $\beta_2$ in the $r_t \leq 0$ regime is in general larger than $\beta_6$, meaning the realized measures will contribute more to the VaR\&ES forecasting when $r_t \leq 0$. Such results are consistent with our expectation and prove the proposed threshold framework can successfully capture the volatility asymmetry.

Third, Realized-Threshold-ES-CARE-RR generates smaller $\sigma_{u}$ values than Realized-Threshold-ES-CARE-RV. This result is consistent with the findings in Martins and van Dijk (2007) and Christensen and Podolskij (2007): RR can have much lower mean squared error than RV, which might provide RR with
higher accuracy and efficiency in volatility estimation and forecasting. The $\sigma_{u} $ of Realized-Threshold-ES-CARE-SSRV is smaller than that of Realized-Threshold-ES-CARE-RV. Comparing Realized-Threshold-ES-CARE-RR and Realized-Threshold-ES-CARE-SSRR, the $\sigma_{u}$ estimates are quite close and are consistent with the findings in Gerlach and Wang (2016b). It seems that $\sigma_{u}$ estimates from the models with ScRV and ScRR are not improved compared with model employing RV and RR. The results here are in general consistent with the subsequently discussed forecasting performance.

Last, as discussed in Martins and van Dijk (2007), RR is biased as a true volatility estimator, if each day $t$ is divided into finite number of equally sized intervals.
However, the RR or SSRR in the Realized-Threshold-ES-CARE models are not required to be unbiased, because the parameters in the
model can adjust such bias: an advantage of using the Realized-GARCH framework which is inherited by the Realized(-Threshold)-ES-CARE frameworks.

\begin{table}[h]
\begin{center}
\caption{\label{para_est_table} \small The estimated Re-Threshold-ES-CARE parameters for the 1st forecasting step with S\&P500.}\tabcolsep=10pt
\tiny
\begin{tabular}{lcccccccc} \hline
Parameters&RV&RR&ScRV&ScRR&SSRV&SSRR\\
\hline
$\beta_1$&-0.2019&-0.1586&-0.2449&-0.2227&-0.2198&-0.1779\\
$\beta_2$&-0.6893&-1.0399&-0.6059&-0.7013&-0.8463&-0.9625\\
$\beta_3$&0.7358&0.7461&0.7223&0.7333&0.7042&0.7315\\
$\beta_4$&-0.0538&-0.0026&-0.0765&-0.0697&-0.0938&-0.0410\\
$\beta_5$&-0.4817&-1.0102&-0.4137&-0.6289&-0.6854&-0.9343\\
$\beta_6$&0.7430&0.6876&0.7465&0.7010&0.6848&0.6735\\
$\tau$&0.001851&0.001873&0.001888&0.002012&0.001957&0.001980\\
$\xi$  &-0.0516&0.0141&-0.1375&-0.0900&-0.0677&-0.0135\\
$\varphi$ &0.3783&0.2410&0.4496&0.3750&0.3441&0.2741\\
$\tau_1$&0.0800&0.0725&0.0855&0.1018&0.1237&0.0962\\
$\tau_2$&0.3653&0.1745&0.3764&0.2342&0.2140&0.1540\\
$\sigma_{u}$&0.2551&0.1504&0.2915&0.2112&0.2037&0.1577\\
\hline
\end{tabular}
\end{center}
\end{table}

\subsection{Tail Risk Forecasting}

$\alpha= 1\%$ is employed for both one day ahead Value-at-Risk (VaR) and Expected Shortfall (ES) forecasting study for the 7 indices.


A rolling window with fixed in-sample size is employed for estimation to produce each one step ahead forecast in the forecast period; the in-sample size $n$ is
given in Table \ref{var_fore_table} for each series, which differs due to non-trading days in each market. In order to see the
performance during the GFC period, the initial date of the forecast sample is chosen as the beginning of 2008. On average,
2111 one day ahead VaR and ES forecasts are generated for each return series from a range of models.

24 models are tested and compared in this section. These include the proposed Realized-ES-CARE and Realized-Threshold-ES-CARE models (estimated with adaptive MCMC) with different input measures of volatility: RV \& RR, scaled RV \& RR and sub-sampled RV \& RR.

The proposed ES-CARE, original ES-CAViaR-Add and ES-CAViaR-Mult models (estimated with adaptive MCMC) are also included in the study. We have also tested the conventional GARCH (Bollerslev, 1986), EGARCH (Nelson, 1991) and GJR-GARCH (Glosten, Jagannathan and Runkle, 1993) with Student-t distribution; the GARCH employing Hansen's skewed-t distribution (Hansen, 1994) and Realized-GARCH with Gaussian and Student-t return equation error distributions (Hansen, Huang and Shek, 2012).

Further, a filtered GARCH-t historical simulation (GARCH-t-HS) approach is also included, where a GARCH-t is fit to the in-sample data. Using all the in-sample data the standardised VaR and ES are estimated via historical simulation, from the sample of returns (e.g. $r_1,\ldots,r_n$ divided by their GARCH-estimated conditional standard deviation (i.e. $r_t/\sqrt{\hat{h_t}}$).
Then final forecasts of VaR, ES are found by multiplying the standardised VaR, ES estimates by the forecast $\sqrt{\hat{h}_{t+1}}$ from the GARCH-t model.

Finally, the Threshold-GARCH model (Li and Li, (1996) and Brooks, (2001)) incorporating Hansen's Skewed-t distribution (T-GARCH-Skew-t) is also tested. All these aforementioned models are estimated by ML, using the Econometrics toolbox in Matlab (GARCH-t, EGARCH-t, GJR-t and GARCH-t-HS) or code developed by the authors (CARE-SAV, T-GARCH-Skew-t and Realized-GARCH). The actual forecast sample sizes $m$, in each series, are given in Table \ref{var_fore_table}.

Firstly, the VaR violation rate is employed to assess the VaR forecasting accuracy. VRate is simply the proportion of returns that exceed the forecasted VaR level in the forecasting period. Models with VRate closest to nominal quantile level $\alpha=1\%$ are preferred.

Several standard quantile accuracy and independence tests are also employed: the unconditional coverage (UC) and conditional
coverage (CC) tests of Kupiec (1995) and Christoffersen (1998) respectively, as well as the dynamic quantile (DQ) test of
Engle and Manganelli (2004) and the VQR test of Gaglione et al. (2011). Finally, the standard quantile loss function is also employed to compare the models for VaR forecast accuracy. Since the standard quantile loss function is strictly consistent, e.g. the expected loss is a minimum at the true quantile series. Thus, the most accurate VaR forecasting model
should produce the minimized quantile loss function, given as:
\begin{equation}\label{q_loss}
\sum_{t=n+1}^{n+m}(\alpha-I(r_t<Q_t))(r_t-Q_t)  \,\, ,
\end{equation}
where $Q_{n+1},\ldots,Q_{n+m}$ is a series of quantile forecasts at level $\alpha$ for the observations $r_{n+1},\ldots,r_{n+m}$.

\subsubsection{\normalsize Value at Risk}

Table \ref{var_fore_table} presents the VRates for each model over the 7 indices. $\alpha = 1\%$ is the target violation rate. For each time series, the models are ranked according to the deviations to the 1\% target rate. Then the average these ranks across all 7 markets is presented in the "Avg Rank" column in Table \ref{var_fore_table}, to compare the overall performance of each model. A box indicates the model with VRate closest to 1\% in each market, while bold indicates the VRate is significantly different to 1\% by the UC test.

As presented in Table \ref{var_fore_table}, overall the best ranked models are proposed ES-CARE and Realized-ES-CARE-SSRR models, followed by the GARCH-Skew-t model. For 5 series the proposed ES-CARE or Realized(-Threshold)-ES-CARE models produce the best or second best VRates. Models, including GARCH-Skew-t, T-GARCH-Skew-t and CARE, also generate quite competitive VRate results. Using quantile loss, now we compare Realized(-Threshold)-ES-CARE type models and these models in more detail with respect to economic efficiency, and provide evidence on why the proposed models are preferred in VaR forecasting.

\begin{table}[hbt!]
\begin{center}
\caption{\label{var_fore_table} \small 1\% VaR Forecasting VRate with different models on 7 indices.}\tabcolsep=10pt
\tiny
\begin{tabular}{lccccccccccc} \hline

Model&S\&P500&NASDAQ&HK&FTSE&DAX&SMI&ASX200&Avg Rank\\ \hline
GARCH-t&\bf{1.467\%}&\bf{1.895\%}&\bf{1.652\%}&\bf{1.731\%}&1.362\%&\bf{1.617\%}&\bf{1.702\%}&20.71\\
EGARCH-t&\bf{1.514\%}&\bf{1.611\%}&1.215\%&\bf{1.777\%}&1.408\%&\bf{1.712\%}&\bf{1.466\%}&19.86\\
GJR-GARCH-t    &\bf{1.467\%}&\bf{1.563\%}&1.263\%&\bf{1.777\%}&1.408\%&\bf{1.759\%}&\bf{1.513\%}&19.57\\
GARCH-t-HS    &1.230\%&\bf{1.563\%}&1.263\%&1.123\%&\cb{1.127\%}&1.284\%&0.898\%&8.29\\
GARCH-Skew-t&\cb{1.088\%}&\cb{1.042\%}&1.263\%&1.169\%&\fbox{0.939\%}&1.331\%&0.804\%&\cb{7.14}\\
T-GARCH-Skew-t&\fbox{0.994\%}&\fbox{0.995\%}&1.312\%&1.356\%&1.174\%&1.331\%&\fbox{1.040\%}&8.00\\
CARE   &1.278\%&\bf{1.563\%}&\fbox{1.020\%}&1.310\%&1.221\%&1.284\%&1.229\%&9.86\\
Re-GARCH-RV-GG   &\bf{2.130\%}&\bf{1.942\%}&\bf{2.818\%}&\bf{1.777\%}&\bf{2.300\%}&\bf{1.807\%}&\bf{1.560\%}&23.57\\
Re-GARCH-RV-tG      &\bf{1.467\%}&1.326\%&\bf{1.992\%}&1.310\%&\bf{1.596\%}&\fbox{1.141\%}&1.229\%&14.29\\
ES-CAViaR-Add&\bf{1.467\%}&\bf{1.516\%}&1.215\%&1.216\%&1.268\%&1.236\%&\cb{0.946\%}&10.57\\
ES-CAViaR-Mult&1.278\%&1.421\%&1.166\%&1.216\%&1.315\%&1.236\%&\cb{0.946\%}&7.71\\
ES-CARE&1.278\%&1.421\%&1.166\%&1.169\%&1.315\%&1.236\%&\cb{0.946\%}&\fbox{7.00}\\
Re-ES-CARE-RV&1.278\%&\bf{1.705\%}&\bf{2.187\%}&1.169\%&1.315\%&1.331\%&0.804\%&13.86\\
Re-ES-CARE-RR&\cb{1.088\%}&1.374\%&1.263\%&0.889\%&1.221\%&1.427\%&0.709\%&9.29\\
Re-ES-CARE-ScRV&1.325\%&\bf{1.658\%}&1.166\%&1.169\%&1.315\%&\cb{1.189\%}&0.898\%&10.00\\
Re-ES-CARE-ScRR&1.278\%&\bf{1.468\%}&\fbox{1.020\%}&1.123\%&1.221\%&1.236\%&0.709\%&7.57\\
Re-ES-CARE-SSRV&1.372\%&\bf{1.468\%}&1.215\%&\cb{1.076\%}&1.315\%&1.284\%&0.709\%&10.71\\
Re-ES-CARE-SSRR&1.278\%&1.374\%&1.166\%&\fbox{0.935\%}&1.174\%&1.331\%&0.709\%&\fbox{7.00}\\
Re-T-ES-CARE-RV&1.183\%&\bf{1.516\%}&\bf{2.041\%}&1.123\%&1.362\%&1.284\%&0.851\%&11.71\\
Re-T-ES-CARE-RR&1.136\%&1.184\%&1.263\%&0.889\%&1.268\%&\bf{1.617\%}&0.662\%&10.86\\
Re-T-ES-CARE-ScRV&1.230\%&1.421\%&1.263\%&1.169\%&1.174\%&1.331\%&0.851\%&8.57\\
Re-T-ES-CARE-ScRR&1.278\%&\bf{1.468\%}&\cb{1.069\%}&1.076\%&1.268\%&1.427\%&0.709\%&9.43\\
Re-T-ES-CARE-SSRV&1.420\%&1.279\%&1.166\%&1.076\%&1.362\%&1.427\%&0.851\%&10.14\\
Re-T-ES-CARE-SSRR&1.372\%&1.137\%&1.166\%&\fbox{0.935\%}&1.315\%&\bf{1.569\%}&0.757\%&9.86\\ \hline
m (Forecasting steps) &2113&2111&2058&2138&2130&2103&2115&2114\\
n (In-sample size)&1905&1892&1890&1943&1936&1930&1871&1916\\

\hline
\end{tabular}
\end{center}
\emph{Note}:\small   Box indicates the favored models based on ESRate, blue shading indicates the 2nd ranked model, for each series and average rank. Bold indicates the violation rate is
significantly different to 1\% by the UC test. $m$ is the out-of-sample size, and $n$ is in-sample size. Re-GARCH stands for the Realized-GARCH
type models. Re-ES-CARE and Re-T-ES-CARE represent Realized-ES-CARE and Realized-T-ES-CARE type models respectively.
\end{table}

The quantile loss results are presented in Table \ref{quanitl_loss_table} for each model for each series. The average rank based on ranks of quantile loss across 7 markets is calculated and shown in the ``Avg Rank'' column. 6 of the 7 series have the lowest loss produced by one of the Realized(-Threshold)-ES-CARE type models (4 from Realized-Threshold-ES-CARE type models and 2 from Realized-ES-CARE type models).  The best average ranked models are Realized-ES-CARE and Realized-Threshold-ES-CARE frameworks employing SSRV. The quantile loss values for the ES-CAViaR-Add and CARE models are on average slightly higher than that of ES-CARE model, meaning ES-CARE model produces more accurate VaR forecasting results. Further, the quantile loss values of Realized-(Threshold)-ES-CARE type models are clearly lower compared to ES-CARE model, which proves that the extra efficiency can be gained by employing the realized measures through the measurement equation.

\begin{table}[hbt!]
\begin{center}
\caption{\label{quanitl_loss_table} \small 1\% VaR Forecasting quantile loss on 7 indices.}\tabcolsep=10pt
\tiny
\begin{tabular}{lccccccccccc} \hline

Model&S\&P500&NASDAQ&HK&FTSE&DAX&SMI&ASX200&Avg Rank\\
\hline
GARCH-t&81.8&92.1&98.4&81.5&93.4&\cred{88.0}&69.7&\cred{19.86}\\
EGARCH-t&80.3&92.2&90.3&76.9&92.2&83.1&67.3&13.29\\
GJR-GARCH-t    &77.6&89.8&92.2&77.9&\cred{93.9}&85.7&67.9&14.43\\
GARCH-t-HS    &81.8&91.5&96.9&80.3&93.9&86.3&69.5&18.71\\
GARCH-Skew-t&81.7&90.9&97.3&80.1&93.6&86.1&69.8&18.29\\
T-GARCH-Skew-t&76.1&87.3&91.3&78.4&91.5&83.8&68.3&10.71\\
CARE   &\bf{84.2}&\bf{95.5}&93.0&\bf{82.7}&93.3&\bf{89.8}&\bf{77.3}&\bf{21.00}\\
Re-GARCH-RV-GG   &80.0&87.3&\bf{119.0}&78.1&\bf{95.2}&83.4&66.1&15.00\\
Re-GARCH-RV-tG      &77.1&\fbox{85.3}&\cred{108.6}&77.0&91.7&82&65.4&10.29\\
ES-CAViaR-Add&\bf{84.2}&\cred{93.5}&94.9&81.1&92.8&86.3&\cred{71.9}&19.57\\
ES-CAViaR-Mult&\cred{83.5}&93.3&95.3&\cred{81.7}&93.2&85.7&71.7&19.29\\
ES-CARE&83.4&93.3&95.3&81.6&93.2&85.7&71.8&19.14\\
Re-ES-CARE-RV&76.1&91.6&106.2&76.5&91.6&81.4&\cb{65.2}&11.14\\
Re-ES-CARE-RR&\cb{73.2}&86.5&101.3&76.3&90.1&79.0&67.4&8.14\\
Re-ES-CARE-ScRV&77.3&89.8&96.5&76.4&93.9&82.9&66.3&12.86\\
Re-ES-CARE-ScRR&76.0&89.4&91.8&76.3&91.0&80.6&67.8&8.43\\
Re-ES-CARE-SSRV&74.8&87.4&91.1&75.8&90.2&\cb{78.7}&66.2&\fbox{5.43}\\
Re-ES-CARE-SSRR&\fbox{72.7}&\cb{86.2}&96.9&75.5&89.7&\fbox{78.6}&66.7&\cb{5.71}\\
Re-T-ES-CARE-RV&78.4&91.5&102.2&\fbox{74.8}&91.9&80.6&\fbox{65.0}&10.29\\
Re-T-ES-CARE-RR&76.6&88.5&98.8&75.5&89.5&80.0&66.2&8.14\\
Re-T-ES-CARE-ScRV&79.9&90.4&92.4&74.9&92.2&82.2&65.8&10.00\\
Re-T-ES-CARE-ScRR&78.5&90.2&\cb{89.3}&\cb{74.9}&90.1&81.6&66.6&8.29\\
Re-T-ES-CARE-SSRV&77.8&88.0&\fbox{88.2}&75.1&\cb{89.2}&80.0&65.9&\fbox{5.43}\\
Re-T-ES-CARE-SSRR&76.3&87.7&94.5&75.7&\fbox{89.1}&79.7&66.5&6.57\\
\hline
\end{tabular}
\end{center}
\emph{Note}:\small  Box indicates the favoured model, blue shading indicates the 2nd ranked model, bold indicates the least favoured model, red shading indicates the 2nd lowest ranked model, in each column.
\end{table}

\begin{figure}[hbt!]
     \centering
\includegraphics[width=.7\textwidth]{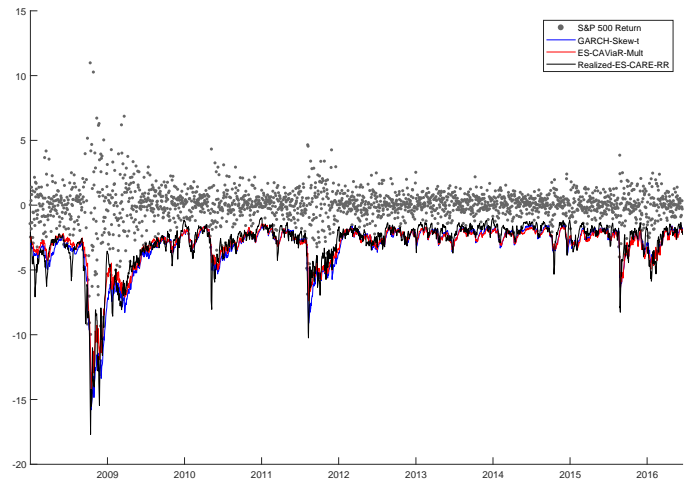}
\caption{\label{var_forecast_fig} S\&P 500 VaR forecasts with GARCH-Skew-t, ES-CAViaR-Mult and Realized-ES-CARE-RR.VRates: 1.088\%, 1.278\% and 1.088\%.}
\end{figure}

\begin{figure}[hbt!]
     \centering
\includegraphics[width=.7\textwidth]{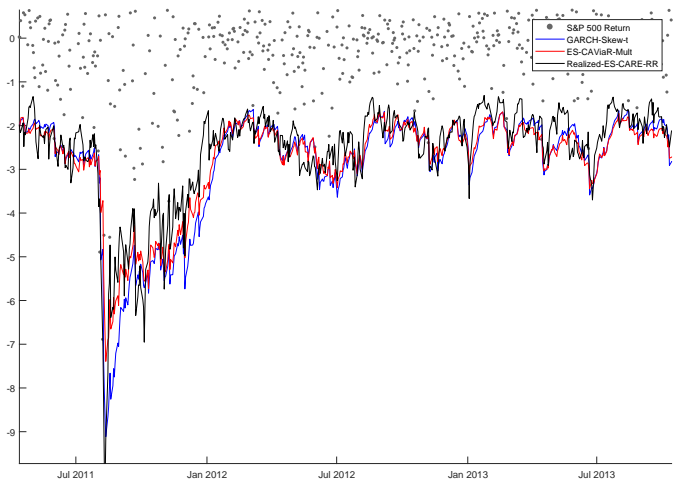}
\caption{\label{var_forecast_fig1} S\&P 500 VaR forecasts (zoomed in) with GARCH-Skew-t, ES-CAViaR-Mult and Realized-ES-CARE-RR. VRates: 1.088\%, 1.278\% and 1.088\%. Quantile loss: 81.7,  83.5 and 73.2.}
\end{figure}

Figure \ref{var_forecast_fig} and \ref{var_forecast_fig1} provide further evidence on how and why the proposed Realized-ES-CARE framework generates clearly lower quantile loss compared with other models, combined
with relatively accurate VRates. Specifically, the VaR violation rates for the GARCH-Skew-t, ES-CAViaR-Mult and Realized-ES-CARE-RR models are 1.088\%, 1.278\% and 1.088\% respectively, for the S\&P500 returns, e.g. the three models are very similar by that metric. However, from Table \ref{quanitl_loss_table}, the quantile loss values for the 3 models are 81.7,  83.5 and 73.2 respectively, meaning the Realized-ES-CARE-RR model is the
most accurate model having clearly the lowest quantile loss. Through close inspection of Figure \ref{var_forecast_fig1}, the GARCH-Skew-t and ES-CAViaR-Mult have VaR forecasts typically quite close together in value, driving their close quantile loss values. However, both these models generate clearly more extreme (in the negative direction) VaR forecasts
on most days in the US market, compared to the Realized-ES-CARE-RR. This means that the capital set aside by financial institutions to cover extreme losses, based on such VaR forecasts, is usually at a higher level for the GARCH-Skew-t or ES-CAViaR-Mult models, than for the Realized-ES-CARE-RR.

In other words, the Realized-ES-CARE-RR model produces VaR forecasts that are relatively close to nominal VRate and are closer to the true VaR series, as measured by the loss function. VaR forecasts of this model are also closer to the data and less extreme, implying that lower amounts of capital are needed to protect against market risk. Given the forecasting steps  $m=2113$ for S\&P 500, the forecasts from Realized-ES-CARE-RR were less extreme than those from ES-CAViaR-Mult on 1396 days (66\%) in the forecast period, which clearly demonstrates the advantageous of employing the RR through the measurement equation.

This suggests a higher level of information (and cost) efficiency regarding risk levels for the Realized-ES-CARE-RR model, likely coming from the improved the model specification and increased statistical efficiency of the realized range series over squared returns, compared to the ES-CAViaR-Mult and GARCH-Skew-t models. Since the economic capital is determined by financial institutions' own model and should be
directly proportional to the VaR forecast, the Realized-ES-CARE-RR model is able to decrease the cost capital allocation and increase the profitability of
these institutions, by freeing up part of the regulatory capital from risk coverage into investment, while still providing sufficient and more than
adequate protection against violations. The more accurate and often less extreme VaR forecasts produced by Realized-ES-CARE-RR are particularly
strategically important to the decision makers in the financial sector. This extra efficiency is also often observed for the Realized(-Threshold)-ES-CARE type
models in the other markets/assets.

Further, during the periods with high volatility including GFC, when there is a persistence of extreme returns, the Realized-ES-CARE-RR VaR forecasts "recover" the fastest among the 3 models, presented through close inspection as in Figure \ref{var_forecast_fig1}, in terms of being marginally the fastest to produce forecasts that again rejoin and follow the tail of the return data. Traditional GARCH models tend to over-react to extreme events and to be subsequently very slow to recover, due to their frequently estimated very high level of persistence, as discussed in Harvey and Chakravarty (2009). Realized(-Threshold)-ES-CARE models clearly improve the performance on this aspect. Generally, the Realized(-Threshold)-ES-CARE models better describe the dynamics in the volatility, compared to the traditional GARCH model and ES-CAViaR-Mult type models, thus largely improving the responsiveness and accuracy of the risk level forecasts, especially after high volatility periods.

Several tests are employed to statistically assess the forecast accuracy and independence of violations from each VaR forecast model.
Table \ref{var_backtest_table} shows the number of return series (out of 7) in which each 1\% VaR forecast model is rejected for each test,
conducted at a 5\% significance level. The Realized(-Threshold)-ES-CARE type models are generally less or equally likely to be rejected by the back tests
compared to other models. The Realized-Threshold-ES-CARE-ScRV achieves the least number of rejections and is rejected once, followed by Gt-HS, T-GARCH-Skew-t, Realized-ES-CARE-RV, Realized-Threshold-ES-CARE-RV and Realized-Threshold-ES-CARE-SSRV models. The G-t and Realized-GARCH-GG are rejected in all 7 series, and the EGARCH-t and GJR-GARCH-t models are rejected in 6 series, respectively.

\begin{table}[hbt!]
\begin{center}
\caption{\label{var_backtest_table} \small Counts of 1\% VaR  rejections with UC, CC, IND, DQ and VQR tests for different models on 7 indices.}\tabcolsep=10pt
\tiny
\begin{tabular}{lccccccccccc} \hline

Model&UC&CC1&DQ1&DQ4&VQR&Total\\
\hline
GARCH-t&6&6&7&5&4&\bf{7}\\
EGARCH-t&5&3&4&5&2&\cred{6}\\
GJR-GARCH-t    &5&3&6&4&3&\cred{6}\\
GARCH-t-HS    &1&1&1&2&0&\cb{2}\\
GARCH-Skew-t&0&0&1&3&0&3\\
T-GARCH-Skew-t&0&0&0&2&0&\cb{2}\\
CARE   &1&1&0&4&0&4\\
Re-GARCH-RV-GG   &7&7&7&7&4&\bf{7}\\
Re-GARCH-RV-tG      &3&2&2&1&3&3\\
ES-CAViaR-Add&2&0&0&3&0&4\\
ES-CAViaR-Mult&0&0&0&3&0&3\\
ES-CARE&0&0&0&3&1&4\\
Re-ES-CARE-RV&2&2&2&2&2&\cb{2}\\
Re-ES-CARE-RR&0&1&1&2&2&3\\
Re-ES-CARE-ScRV&1&1&1&2&2&3\\
Re-ES-CARE-ScRR&1&1&2&3&0&3\\
Re-ES-CARE-SSRV&1&1&2&4&2&4\\
Re-ES-CARE-SSRR&0&1&1&2&2&3\\
Re-T-ES-CARE-RV&2&1&1&1&1&\cb{2}\\
Re-T-ES-CARE-RR&1&1&2&2&2&3\\
Re-T-ES-CARE-ScRV&0&0&0&1&0&\fbox{1}\\
Re-T-ES-CARE-ScRR&1&1&2&2&0&3\\
Re-T-ES-CARE-SSRV&0&1&1&2&0&\cb{2}\\
Re-T-ES-CARE-SSRR&1&1&2&2&1&3\\
\hline
\end{tabular}
\end{center}
\emph{Note}:\small Box indicates the model with least number of rejections, blue shading indicates the model with 2nd least number of rejections, bold indicates the model with the highest number of rejections, red shading indicates the model 2nd highest number of rejections.
All tests are conducted at 5\% significance level.
\end{table}

\subsection{\normalsize Expected Shortfall and Expectile Level}
\begin{figure}[htp]
     \centering
\includegraphics[width=.9\textwidth]{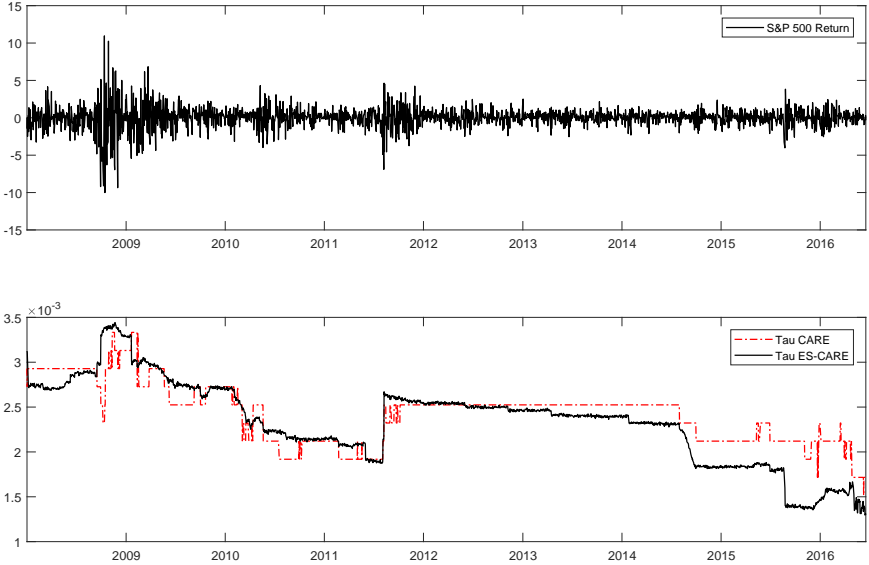}
\caption{\label{tau_comparision figure} For S\&P 500 forecasting, top plot visualizes S\&P 500 returns, and bottom plot visualizes $m=2113$ estimated $\tau$ parameters from CARE and ES-CARE models.}
\end{figure}

The same 24 models are employed to generate one step ahead forecasts of 1\% ES for all 7 series during the forecasting period. Before checking the ES forecasting results, Figure \ref{tau_comparision figure} visualizes the 2113 estimated expectile level $\tau$ parameters from ES-CARE and CARE models during the forecasting period of S\&P 500. The $\tau$ of CARE is selected based on the grid search and violation rate approach of Taylor (2008), as discussed in Section \ref{model_review_section}.

Although in general the $\tau$ values estimated by ES-CARE and CARE are close to each other, we can clearly see the ES-CARE model produces more dynamic $\tau$ values, e.g. during the 2008 GFC and the mid-2012 to 2014 period. Especially, the estimated $\tau$ from ES-CARE is much more responsive to the volatility jump. Such improved responsiveness will potentially improve the accuracy of tail risk forecasts. As discussed in Section \ref{model_section}, the selection procedure of $\tau$ with the ES-CARE model is based on a strictly consistent VaR\&ES joint loss function, unlike the grid search approach. More results will be provided in the next section to support the improved ES forecasting performance from ES-CARE compared to CARE.

In addition, during the low volatility period, e.g. mid-2014 to mid-2015, the estimated $\tau$ values from ES-CARE model are clearly smaller than those from CARE. Based on Equation (\ref{expectile_es_equation}), the ratio between ES and VaR is linearly proportional to $\tau$. Therefore, the ES to VaR ratios produced by ES-CARE model, during mid-2014 to mid-2015 low volatility period (less cases of extreme returns), are smaller than that produced by CARE. Such results are consistent with the definition of VaR and ES, and lend some evidence on why the VaR and ES forecasts generated by ES-CARE are more efficient than CARE, to be further discussed in the following section.

\begin{figure}[htp]
     \centering
\includegraphics[width=.9\textwidth]{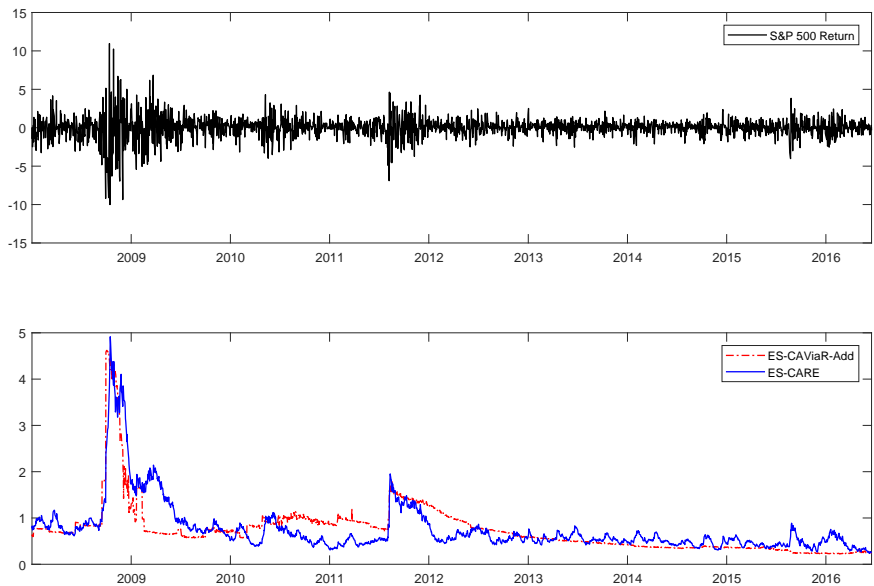}
\caption{\label{xt_comparision figure} For S\&P 500 forecasting, top plot visualizes S\&P 500 returns, and bottom plot visualizes $m=2113$ VaR and ES forecasts differences from ES-CAViaR-Add and ES-CARE models.}
\end{figure}

Further, as discussed in Section \ref{model_review_section}, the ES component in the ES-CAViaR-Add framework would produce the in-sample $w_{t}$ as presented in Figure \ref{wt_example}, resulting ES dynamics that could be potentially improved. Figure \ref{xt_comparision figure} visualize $m=2113$ VaR and ES forecasts differences from ES-CAViaR-Add and ES-CARE models. Clearly, more dynamic ES and VaR differences are produced by the ES-CARE which incorporates a more flexible ES regression component compared with ES-CAViaR-Add. More specifically, the difference between VaR and ES should be larger during the high volatility period, based on the definition of ES, e.g. as illustrated in the 2008 GFC period. However, taking the period of early-2009 to mid-2009 as example, apparently the ES-CARE creates more responsive ES forecasting results than ES-CAViaR-Add. In the following Section, we will quantify the improvement of VaR\& ES forecasting results from ES-CARE model compared with ES-CAViaR-Add.

\subsubsection{\normalsize VaR\&ES Joint Loss Function}

In this section, the joint VaR\&ES loss function study is conducted to compare the models VaR and ES forecasts jointly, and to help clarify and quantify any
extra efficiency can be gained from the Realized(-Threshold)-ES-CARE ES forecasts compared to its competitors.

Fissler and Ziegel (2016) developed a family of loss functions that are a joint function of the associated VaR and ES series. This loss function family are
 strictly consistent for the true VaR and ES series, i.e. they are uniquely minimized by the true VaR and ES series. The general function family form is:
\begin{eqnarray*}
S_t(r_t, VaR_t, ES_t) &=& (I_t -\alpha)G_1(VaR_t) - I_tG_1(r_t) +  G_2(ES_t)\left(ES_t-VaR_t + \frac{I_t}{\alpha}(VaR_t-r_t)\right) \\
                      &-& H(ES_t) + a(r_t) \, ,
\end{eqnarray*}
where $I_t=1$ if $r_t<VaR_t$ and 0 otherwise for $t=1,\ldots,T$, $G_1()$ is increasing, $G_2()$ is strictly increasing and strictly convex,
$G_2 = H^{'}$ and $\lim_{x\to -\infty} G_2(x) = 0$ and $a(\cdot)$ is a real-valued integrable function.

As discussed in Taylor (2017),  making the choices: $G_1(x) =0$,
$G_2(x) = -1/x$, $H(x)= -\text{log}(-x)$ and  $a= 1-\text{log} (1-\alpha)$, which satisfy the required criteria, returns the
scoring function (defined $r_t$ to have zero mean):
\begin{eqnarray}\label{es_caviar_log_score}
S_t(r_t, VaR_t, ES_t) = -\text{log} \left( \frac{\alpha-1}{\text{ES}_t} \right) - {\frac{(r_t-Q_t)(\alpha-I(r_t\leq Q_t))}{\alpha \text{ES}_t}},
\end{eqnarray}

where the loss function is $S = \sum_{t-1}^T S_t$. Taylor (2017) referred expression (\ref{es_caviar_log_score}) as AL log score. Compared with the likelihood function as in Equation (\ref{es_caviar_like_equation}), Equation (\ref{es_caviar_log_score}) is exactly the negative of the AL log-likelihood, and is a strictly consistent scoring rule that is jointly minimized by the true VaR and ES series. We use this to informally and jointly assess and compare the VaR and ES forecasts from all models.

Tables \ref{veloss} shows the loss function values $S$, calculated using Equation (\ref{es_caviar_log_score}), which jointly assesses the
accuracy of each model's VaR and ES forecasting series, during the forecast period for each market. Generally, the Realized(-Threshold)-ES-CARE models are better ranked with lower loss than other models in most markets. For all 7 markets, the best ranked models are from Realized(-Threshold)-ES-CARE families, with Realized-Threshold-ES-CARE employing SSRV and SSRR achieving the best performance overall.

In addition, overall the VaR \& ES joint loss values from Realized-Threshold-ES-CARE specifications are consistently lower than that from Realized-ES-CARE specifications, under the same choice of realized measures (except RR). For example, Realized-Threshold-ES-CARE-SSRR is better ranked than Realized-ES-CARE-SSRR. This proves the validity of the threshold specifications for the VaR to ES dynamics.

Lastly, overall the ES-CARE model is better ranked than ES-CAViaR-Add, ES-CAViaR-Mult and CARE models. This again proves the validity of the proposed ES-CARE framework, even without incorporating the realized measures.

\begin{table}[hbt!]
\begin{center}
\caption{\label{veloss} \small VaR and ES joint loss function values across the markets; $\alpha=1\%$.}\tabcolsep=10pt
\tiny
\begin{tabular}{lccccccccccccc} \hline

Model&S\&P500&NASDAQ&HK&FTSE&DAX&SMI&ASX200&Avg Rank\\
\hline
GARCH-t&4795.0&5067.2&5144.4&4872.1&\cred{5285.2}&\cred{4987.1}&4531.9&\bf{20.57}\\
EGARCH-t&4800.9&5068.7&4985.2&4837.6&5277.3&4905.2&4503.7&17.71\\
GJR-GARCH-t    &4665.5&4967.7&5009.9&4793.8&\bf{5315.8}&\bf{4993.5}&4475.2&16.29\\
GARCH-t-HS    &4768.8&5031.3&5100.9&4811.8&5274.5&4884.3&4510.4&18.00\\
GARCH-Skew-t&4758.9&5008.6&5104.5&4814.6&5254.1&4882.7&4508.8&17.57\\
T-GARCH-Skew-t&4613.0&4909.4&4966.4&4800.4&5195.9&4873.3&4472.7&10.86\\
CARE   &\cred{4836.7}&\bf{5201.5}&5018.5&\bf{4890.9}&5231.8&4973.2&\bf{4793.4}&\cred{20.29}\\
Re-GARCH-RV-GG   &4706.0&4948.3&\bf{5673.1}&4768.5&5275.2&4878.0&4432.7&15.71\\
Re-GARCH-RV-tG      &4590.8&\cb{4875.4}&5288.3&4706.4&5146.0&4778.0&4386.4&10.14\\
ES-CAViaR-Add&\bf{4844.0}&\cred{5099.7}&5069.0&4859.6&5237.1&4941.2&\cred{4596.9}&20.00\\
ES-CAViaR-Mult&4833.6&5068.2&5071.5&\cred{4875.6}&5234.8&4909.0&4589.5&19.71\\
ES-CARE&4832.6&5068.1&5069.8&4874.0&5234.2&4907.7&4591.2&18.86\\
Re-ES-CARE-RV&4551.6&4983.2&\cred{5295.9}&4680.9&5152.0&4757.2&\cred{4380.0}&11.29\\
Re-ES-CARE-RR&\cb{4477.6}&4877.9&5149.3&4667.8&5094.6&4718.2&4459.6&7.86\\
Re-ES-CARE-ScRV&4599.6&4963.2&5090.2&4676.8&5210.8&4759.2&4411.0&11.71\\
Re-ES-CARE-ScRR&4548.3&4934.1&4969.9&4675.5&5120.1&\cb{4705.9}&4480.1&7.71\\
Re-ES-CARE-SSRV&4505.1&4897.6&4974.7&4672.1&5104.2&\fbox{4702.2}&4415.3&5.29\\
Re-ES-CARE-SSRR&\fbox{4455.9}&\fbox{4872.1}&5071.0&4656.4&5089.9&4715.3&4436.2&5.71\\
Re-T-ES-CARE-RV&4615.3&4999.2&5241.7&\cb{4645.1}&5140.5&4753.5&\fbox{4376.1}&10.14\\
Re-T-ES-CARE-RR&4555.9&4926.2&5082.3&4655.4&5077.7&4745.1&4420.0&8.00\\
Re-T-ES-CARE-ScRV&4678.2&4993.3&4995.2&\fbox{4644.8}&5159.3&4748.5&4410.4&9.29\\
Re-T-ES-CARE-ScRR&4626.0&4958.5&\cb{4923.0}&4646.2&5100.4&4727.3&4449&7.43\\
Re-T-ES-CARE-SSRV&4591.3&4920.4&\fbox{4919.9}&4652.2&\cb{5071.7}&4731.0&4405.0&\fbox{4.86}\\
Re-T-ES-CARE-SSRR&4542.6&4914.1&5008.8&4650.9&\fbox{5070.9}&4739.9&4408.5&\cb{5.00}\\
\hline
\end{tabular}
\end{center}
\emph{Note}:\small  Box indicates the favoured model, blue shading indicates the 2nd ranked model, bold indicates the least favoured model,
red shading indicates the 2nd lowest ranked model, in each column.
\end{table}

\begin{figure}[hbt!]
     \centering
\includegraphics[width=.7\textwidth]{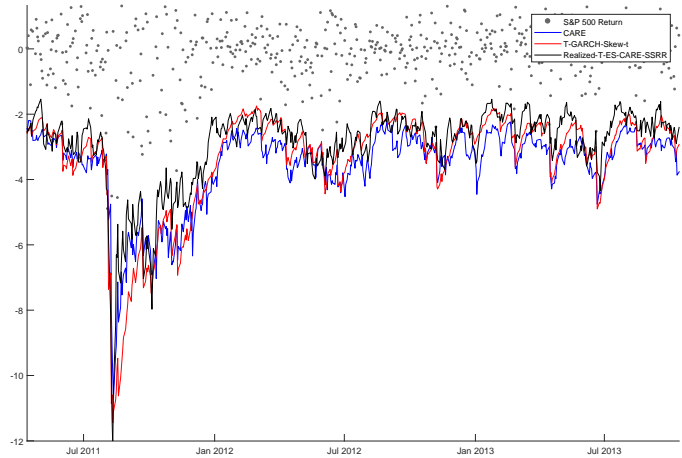}
\caption{\label{Fig_es_fore_zoom_in} S\&P 500 ES forecasts (zoomed in) with CARE, Threshold-GARCH-Skew-t and Realized-Threshold-ES-CARE-SSRR. ESRates: 0.284\%, 0.426\% and 0.473\%. VaR\&ES join loss: 4,836.7, 4613.0, and 4542.6.}
\end{figure}

To further demonstrate the extra forecasting efficiently can be gained by employing the proposed models, Figure \ref{Fig_es_fore_zoom_in} visualizes the ES forecasts from CARE, Threshold-GARCH-Skew-t and Realized-Threshold-ES-CARE-SSRR. Specifically, the ES violation rate of these three models are: 0.284\%, 0.426\% and 0.473\% respectively,
for S\&P500. As studied in Gerlach and Chen (2016, Table 1), the nominal ES quantile levels only have very small variations across different distributions or different degrees of freedom for a Student-t or skewed-t. In general, for 1\% ES the quantile levels that ES to fall is between 0.34\% and 0.37\%. Based on the models with Student-t errors, the implied quantile level that the 1\% ES is estimated to fall at is $\approx$ 0.35\% which is used as the target ES violation rate for the semi-parametric models. If this nominal level is accurate, then the
CARE has the conservative ES violation rate, while the Threshold-GARCH-Skew-t and Realized-Threshold-ES-CARE-SSRR model are slightly anti-conservative (neither are significantly
different to 0.35\% by the UC test).

However, through closer inspection of Figure \ref{Fig_es_fore_zoom_in}, the cost efficiency gains from the Realized-Threshold-ES-CARE-SSRR model are again observed, in a
similar and even clearer pattern to that from the VaR forecasting study. The CARE model achieves a lower than nominal ES VRate by generating relatively more extreme ES forecasts
than the Realized-Threshold-ES-CARE-SSRR model's on 1654 days (78\%). In addition, the ES forecasts from Threshold-GARCH-Skew-t model are more extreme than Realized-Threshold-ES-CARE-SSRR on 1471 days (70\%).

Therefore, the Realized-Threshold-ES-CARE-SSRR  can improve the forecast efficiency and lead to lower capital allocations to protect against extreme returns, compared with the CARE and Threshold-GARCH-Skew-t, while still achieving an accurate violation rate. Again, such extra efficiency is also frequently observed for the Realized(-Threshold)-ES-CARE type models in other time series.

Here, compared with the VaR forecasting, we would like to emphasize the extra efficiency produced by the Realized(-Threshold)-ES-CARE models are more prominent, compared with the original ES-CAViaR and conventional GARCH models. The results lend evidence on fact that the newly developed Realized(-Threshold)-ES-CARE frameworks can produce more accurate and efficient VaR and ES forecasts.

\subsubsection{\normalsize Model Confidence Set}

The model confidence set (MCS) (Hansen, Lunde and Nason, 2011) is utilized to statistically compared the tested models via the VaR and ES joint loss function.

A MCS is a set of models that is constructed such that it will contain the best model with a given level of confidence (90\% is used in our paper). The Matlab code for MCS testing was
downloaded from "www.kevinsheppard.com/MFE\_Toolbox". We adapted code to incorporate the VaR and ES joint loss function values
(Equation (\ref{es_caviar_log_score})) as the loss function during the MCS calculation. The R method which uses absolute values sum during the calculation of test statistic is employed in our paper, details as in page 465 of Hansen, Lunde and Nason (2011).

Table \ref{mcs_r} presents the 90\% MCS using the R methods. Column "Total" counts the total number of times that a model is included in the 90\% MCS across the 7 return series.

Via the R method, 4 Realized(-Threshold)-ES-CARE models, including Realized-ES-CARE-SSRV, Realized-ES-CARE-SSRR, Realized-Threshold-ES-CARE-RR and Realized-Threshold-ES-CARE-SSRR, are included in the MCS for all 7 markets. There are other 4 Realized(-Threshold)-ES-CARE models included in the MCS for 6 times, together with EGARCH-t and Threshold-GARCH-Skew-t.
The GARCH-t and GARCH-Skew-t are only included the MCS for 3 times respectively.

\begin{table}[hbt!]
\begin{center}
\caption{\label{mcs_r} \small 90\% model confidence set results summary with R method, $\alpha=1\%$.}\tabcolsep=10pt
\tiny
\begin{tabular}{lccccccccccccc} \hline
Model&S\&P500&NASDAQ&HK&FTSE&DAX&SMI&ASX200&Total\\
\hline
GARCH-t&0&1&0&1&0&1&0&\bf{3}\\
EGARCH-t&0&1&1&1&1&1&1&\cb{6}\\
GJR-GARCH-t    &0&1&1&1&0&1&1&5\\
GARCH-t-HS    &0&1&0&1&0&1&0&\bf{3}\\
GARCH-Skew-t&0&1&0&1&1&1&0&\cred{4}\\
T-GARCH-Skew-t&0&1&1&1&1&1&1&\cb{6}\\
CARE   &0&0&1&1&1&1&0&\cred{4}\\
Re-GARCH-RV-GG   &0&1&0&1&0&1&1&\cred{4}\\
Re-GARCH-RV-tG      &0&1&0&1&1&1&1&5\\
ES-CAViaR-Add&0&0&1&1&1&1&0&\cred{4}\\
ES-CAViaR-Mult&0&1&1&1&1&1&0&5\\
ES-CARE&0&1&1&1&1&1&0&5\\
Re-ES-CARE-RV&0&1&0&1&1&1&1&5\\
Re-ES-CARE-RR&1&1&0&1&1&1&0&5\\
Re-ES-CARE-ScRV&0&1&0&1&0&1&1&\cred{4}\\
Re-ES-CARE-ScRR&1&1&1&1&1&1&0&\cb{6}\\
Re-ES-CARE-SSRV&1&1&1&1&1&1&1&\fbox{7}\\
Re-ES-CARE-SSRR&1&1&1&1&1&1&1&\fbox{7}\\
Re-T-ES-CARE-RV&0&1&0&1&1&1&1&5\\
Re-T-ES-CARE-RR&1&1&1&1&1&1&1&\fbox{7}\\
Re-T-ES-CARE-ScRV&0&1&1&1&1&1&1&\cb{6}\\
Re-T-ES-CARE-ScRR&0&1&1&1&1&1&1&\cb{6}\\
Re-T-ES-CARE-SSRV&0&1&1&1&1&1&1&\cb{6}\\
Re-T-ES-CARE-SSRR&1&1&1&1&1&1&1&\fbox{7}\\
\hline
\end{tabular}
\end{center}
\emph{Note}:\small Boxes indicate the favoured model, blue shading indicates the 2nd ranked model, bold indicates the least favoured model,
red shading indicates the 2nd lowest ranked model, based on total number of included in the MCS across the 7 markets, higher is better.
\end{table}

{\centering
\section{\normalsize CONCLUSION}\label{conclusion_section}
\par
}
\noindent
In this paper, we propose a realized joint conditional autoregressive expectile and expected shortfall framework which is further extended through incorporating nonlinear specifications. Improvements in the out-of-sample forecasting of tail risk measures are observed, compared to Realized-GARCH model employing realized volatility, and
traditional GARCH and CARE models, as well as the original ES-CAViaR models. Specifically, Realized(-Threshold)-ES-CARE frameworks employing sub-sampled RV and sub-sampled RR generate the best VaR and ES forecasting results in the empirical study of 7 financial return series. With respect to the back testing of VaR forecasts, the Realized(-Threshold)-ES-CARE type models are also generally less likely to be rejected than their counterparts.
Further, the model confidence set results also apparently favour the proposed Realized(-Threshold)-ES-CARE frameworks.
In addition to being more accurate, the Realized(-Threshold)-ES-CARE models generated less extreme tail risk forecasts, regularly allowing smaller amounts of capital allocation without being anti-conservative or significantly inaccurate. Further, even without incorporating the realized measures, the ES-CARE model is still favourable compared with CARE and ES-CAViaR models under almost all the measures and tests considered.

To conclude, the Realized(-Threshold)-ES-CARE type models, especially the ones use sub-sampled RV and sub-sampled RR, should be considered for financial applications when
forecasting tail risk, and should allow financial institutions to more accurately allocate capital under the Basel Capital Accords,
to protect their investments from extreme market movements. This work could be extended by using alternative frequencies of observation for the realized measures; by extending the framework to allow multiple realized measures to appear simultaneously in the model, etc.

\clearpage
\section*{References}
\addcontentsline{toc}{section}{References}
\begin{description}

\item Aigner, D.J. ,Amemiya, T., and Poirier, D. J. (1976). On the Estimation of Production
Frontiers: Maximum Likelihood Estimation of the Parameters of a Discontinuous Density
Function. \emph{International Economic Review}, 17, 377-396.

\item Andersen, T. G. and Bollerslev, T. (1998). Answering the skeptics: Yes, standard volatility models do provide accurate
forecasts. \emph{International economic review}, 885-905.

\item Andersen, T. G., Bollerslev, T., Diebold, F. X. and Labys, P. (2003). Modeling and forecasting realized volatility.
    \emph{Econometrica}, 71(2), 579-625.

\item Artzner, P., Delbaen, F., Eber, J.M., and Heath, D. (1997). Thinking coherently.  \emph{Risk}, 10, 68-71.

\item Artzener, P., Delbaen, F., Eber, J.M., and Heath, D. (1999). Coherent measures of risk.  \emph{Mathematical Finance}, 9, 203-228.

\item Avdulaj, K. and Barunik, J. (2017). A semiparametric nonlinear quantile regression model for financial returns. \emph{Studies in Nonlinear Dynamics \& Econometrics}, 21(1), 81-97.

\item Bollerslev, T. (1986). Generalized Autoregressive Conditional Heteroskedasticity. \emph{Journal of Econometrics}, 31, 307-327.

\item Brooks, C. (2001). A Double‐threshold GARCH Model for the French Franc/Deutschmark exchange rate. \emph{Journal of Forecasting}, 20(2), 135-143.

\item Chen, W., Peters, G., Gerlach, R. and Sisson, S. (2017). Dynamic Quantile Function Models. arXiv:1707.02587.

\item Christensen, K. and Podolskij, M. (2007). Realized range-based estimation of integrated variance. \emph{Journal of Econometrics},
141(2), 323-349.

\item Christoffersen, P. (1998). Evaluating interval forecasts. \emph{International Economic Review}, 39, 841-862.

\item Clements, M.P., Galv\~{a}o, A.B. and Kim, J.H. (2008). Quantile forecasts of daily exchange rate returns from forecasts of realized volatility. \emph{Journal of Empirical Finance}, 15(4), 729-750.

\item Creal, D., Koopman, S.J. and Lucas, A. (2013). Generalized autoregressive score models with applications. \emph{Journal of Applied Econometrics}, 28(5), 777-795.

\item Embrechts, P., Resnick, S.I. and Samorodnitsky, G. (1999). Extreme value theory as a risk management tool. \emph{North American Actuarial Journal}, 3(2), pp.30-41.

\item Engle, R. F. (1982), Autoregressive Conditional Heteroskedasticity with Estimates of the Variance of United Kingdom
Inflations. \emph{Econometrica}, 50, 987-1007.

\item Engle, R. F. and Manganelli, S. (2004). CAViaR: Conditional Autoregressive Value at Risk
by Regression Quantiles. \emph{Journal of Business and Economic Statistics}, 22, 367-381.

\item Fissler, T. and Ziegel, J. F. (2016). Higher order elicibility and Osband's principle. \emph{Annals of Statistics}, in press.

\item Gaglianone, W. P., Lima, L. R., Linton, O. and Smith, D. R. (2011). Evaluating Value-
at-Risk models via quantile regression. \emph{Journal of Business and Economic Statistics},
29, 150-160.

\item Garman, M. B. and Klass, M. J. (1980). On the Estimation of Security Price Volatilities from historical data.
\emph{The Journal of Business}, 67-78.

\item Gelman, A., Carlin, J.B., Stern, H.S. and Rubin, D.B. (2014). \emph{Bayesian data analysis (Vol. 2)}. Boca Raton, FL: CRC press.

\item Gerlach, R, Chen, C.W.S. and Chan, N.Y. (2011). Bayesian time-varying quantile forecasting for value-at-risk in financial markets. \emph{Journal of Business \& Economic Statistics}, 29(4), 481-492.

\item Gerlach, R. and Chen, C.W.S. (2016). Bayesian Expected Shortfall Forecasting Incorporating the Intraday Range,
\emph{Journal of Financial Econometrics}, 14(1), 128-158.

\item Gerlach, R. and Wang, C. (2016a).  Forecasting risk via realized GARCH, incorporating the realized range.
\emph{Quantitative Finance}, 16:4, 501-511.

\item Gerlach, R. and Wang, C. (2016b). Bayesian Semi-parametric Realized Conditional Autoregressive Expectile Models for Tail Risk Forecasting. arXiv preprint arXiv:1612.08488.

\item Gerlach, R., Walpole, D. and Wang, C. (2017). Semi-parametric Bayesian Tail Risk Forecasting Incorporating Realized Measures of Volatility,
\emph{Quantitative Finance}, 17:2, 199-215.

\item Gilli, M. and Kellezi, E. (2006). An Application of Extreme Value Theory for Measuring Financial Risk. \emph{Computational Economics}, 27, 1–23.

\item Giot, P. and Laurent, S. (2004). Modelling daily value-at-risk using realized volatility and ARCH type models. \emph{Journal of Empirical Finance}, 11(3), 379-398.

\item Glosten, L.R., Jagannathan, R. and Runkle, D.E. (1993). On the relation between the expected value and the volatility of the nominal excess return on stocks. \emph{The journal of finance}, 48(5), 1779-1801.

\item Hansen, P. R., Huang, Z. and Shek, H. H. (2012). Realized GARCH: a joint model for returns and realized measures of volatility. \emph{Journal of Applied Econometrics}, 27(6), 877-906.

\item Hansen, P.R., Lunde, A. and Nason, J.M. (2011). The model confidence set. \emph{Econometrica}, 79(2), 453-497.

\item Harvey, A.C. and Chakravarty, T. (2009). Beta-t-EGARCH. Working paper. Earlier version appeared in 2008 as a Cambridge Working paper in Economics, CWPE 0840.

\item Harvey, A.C. (2013). Dynamic Models for Volatility and Heavy Tails, Econometric Society Monograph 52, Cambridge University Press, Cambridge.

\item Koenker, R. and Machado, J.A. (1999). Goodness of fit and related inference processes for quantile regression. \emph{Journal of the american statistical association}, 94(448), 1296-1310.

\item Kupiec, P. H. (1995). Techniques for Verifying the Accuracy of Risk Measurement Models. \emph{The Journal of Derivatives}, 3, 73-84.

\item Li, C. W., and Li, W. K. (1996). On a Double-Threshold Autoregressive Heteroscedastic Time Series Model. \emph{Journal of Applied Econometrics}, 11(3), 253–274.

\item Brooks, C. (2001). A Double-Threshold GARCH Model for the French Franc/Deutschmark Exchange Rate. \emph{Journal of Forecasting}, 20, 135–143.

\item Martens, M. and van Dijk, D. (2007). Measuring volatility with the realized range. \emph{Journal of Econometrics}, 138(1), 181-207.

\item McNeil, A. J. and Frey, R. (2000). Estimation of Tail-Related Risk Measures for Heteroscedastic Financial Time Series: An Extreme Value Approach. \emph{Journal of Empirical Finance} 7, 271–300.

\item Metropolis, N., Rosenbluth, A. W., Rosenbluth, M. N., Teller, A. H., and Teller, E. (1953). Equation of State Calculations by Fast
Computing Machines. \emph{J. Chem. Phys}, 21, 1087-1092.

\item Parkinson, M. (1980). The extreme value method for estimating the variance of the rate of return. \emph{Journal of Business},
53(1), 61.

\item Patton, A.J., Ziegel, J.F. and Chen, R. (2017). Dynamic semiparametric models for expected shortfall (and value-at-risk). arXiv preprint arXiv:1707.05108.

\item Roberts, G. O., Gelman, A. and Gilks, W. R. (1997). Weak convergence and optimal scaling of random walk Metropolis algorithms.
    \emph{The annals of applied probability}, 7(1), 110-120.

\item Taylor, J. (2008). Estimating Value at Risk and Expected Shortfall Using Expectiles. \emph{Journal of Financial Econometrics}, 6,
    231-252.

 \item   Taylor, J. (2017). Forecasting Value at Risk and Expected Shortfall Using a Semiparametric Approach Based on the Asymmetric Laplace Distribution. \emph{Journal of Business and Economic Statistics}, DOI:10.1080/07350015.2017.1281815.

\item Watanabe, T. (2012). Quantile Forecasts of Financial Returns Using Realized GARCH Models. \emph{Japanese Economic Review}, 63(1),
    68-80.

\item Zhang, L., Mykland, P. A., and A\"{i}t-Sahalia, Y. (2005). A tale of two time scales.  \emph{Journal of the American Statistical
    Association}, 100(472).

\item \v{Z}ike\v{s}, F. and Barun\'{i}k, J. (2014). Semi-parametric conditional quantile models for financial returns and realized volatility. \emph{Journal of Financial Econometrics}, 14(1), 185-226.

\end{description}

\end{document}